\documentclass[journal]{vgtc-no-name}         

\ifpdf                                
  \pdfoutput=1\relax                   
  \usepackage{graphicx}                
  \DeclareGraphicsExtensions{.pdf,.png,.jpg,.jpeg} 
  \else                                 
  \usepackage{graphicx}                
  \DeclareGraphicsExtensions{.eps}     
\fi

\graphicspath{{figs/}{./}} 

\usepackage{microtype}                 
\PassOptionsToPackage{warn}{textcomp}  
\usepackage{textcomp}                  
\usepackage{mathptmx}                  
\usepackage{times}                     
         
\usepackage{cite}                      
\usepackage{tabu}                      
\usepackage{booktabs}       
\usepackage{multirow}

\usepackage{amssymb}
\usepackage{amsmath}
\usepackage{amsthm} 
\usepackage{amsfonts}
\usepackage[ruled]{algorithm2e}
\usepackage{tikz}
\usetikzlibrary{arrows}

\usepackage{color}

\newcommand {\mm}[1]        {\ifmmode{#1}\else{\mbox{\(#1\)}}\fi}

\newcommand{\onevec}[1]     {\mm{\mathbf{1}_{#1}}}

\newcommand{\Rspace}        {\mm{\mathbb{R}}}
\newcommand{\Hspace}        {\mm{\mathbb{H}}}
\newcommand{\Mspace}        {\mm{\mathbb{M}}}

\newcommand{\Scal}        {\mm{\mathcal{S}}}
\newcommand{\Tcal}        {\mm{\mathcal{T}}}
\newcommand{\Gcal}        {\mm{\mathcal{G}}}
\newcommand{\Ncal}        {\mm{\mathcal{N}}}

\newcommand{\Ccal}        {\mm{\mathcal{C}}}

\newcommand{\IFS}        {\emph{IFS}}
\newcommand{\LSS}        {\emph{LSS}}
\newcommand{\NMF}        {\emph{NMF}}

\newcommand{\ie}            {{\it i.e.}}
\newcommand{\wrt}            {{\it w.r.t.}}
\newcommand{\eg}            {{\it e.g.}}
\newcommand{\etal}            {{\it et al.}}

\DeclareMathOperator{\distortion}  {\mm{\mathcal{E}}}
\DeclareMathOperator{\discrepancy}  {\mm{\mathcal{D}}}
\DeclareMathOperator*{\argmin}{argmin}

\newcommand{\para}[1]{\vspace{1mm}\noindent{\textbf{#1.~}}}

\newcommand{\pinv}          {{+}}

\usepackage{mathptmx}
\DeclareMathAlphabet{\mathcal}{OMS}{cmsy}{m}{n}
\usepackage{mathtools}
\DeclareMathDelimiter{(}{\mathopen} {operators}{"28}{largesymbols}{"00}
\DeclareMathDelimiter{)}{\mathclose}{operators}{"29}{largesymbols}{"01}

\newcommand{\denselist}{\vspace{-5pt} \itemsep -2pt\parsep=-1pt\partopsep -2pt}

\onlineid{0}

\vgtccategory{Research}

\title{Sketching Merge Trees for Scientific Data Visualization}

\author{Mingzhe Li, Sourabh Palande, Lin Yan, Bei Wang}
\authorfooter{
\item
 Mingzhe Li, Lin Yan, and Bei Wang are with University of Utah. 
 E-mails: mingzhel@cs.utah.edu, lin.yan@utah.edu, beiwang@sci.utah.edu.
\item
Sourabh Palande is with Michigan State University.
E-mail: palandes@msu.edu.
}

\shortauthortitle{Li \MakeLowercase{\textit{et al.}}: Sketching Merge Trees}

\abstract{
Merge trees are a type of topological descriptors that record the connectivity among the sublevel sets of scalar fields. 
They are among the most widely used topological tools in visualization. 
In this paper, we are interested in sketching a set of merge trees. That is, given a large set $\Tcal$ of merge trees, we would like to find a much smaller basis set $\Scal$ such that each tree in $\Tcal$ can be approximately reconstructed from a linear combination of merge trees in $\Scal$. 
A set of high-dimensional vectors can be sketched via matrix sketching techniques such as principal component analysis and column subset selection. 
However, up until now, topological descriptors such as merge trees have not been known to be \emph{sketchable}.  
We develop a framework for sketching a set of merge trees that combines the Gromov-Wasserstein probabilistic matching with techniques from matrix sketching. 
We demonstrate the applications of our framework in sketching merge trees that arise from time-varying scientific simulations. 
Specifically, our framework obtains a much smaller representation of a large set of merge trees for downstream analysis and visualization. It is shown to be useful in identifying good  representatives and outliers with respect to a chosen basis.
Finally, our work shows a promising direction of utilizing randomized linear algebra within scientific visualization.

}


\teaser{
  \centering
  \includegraphics[width=0.8\linewidth]{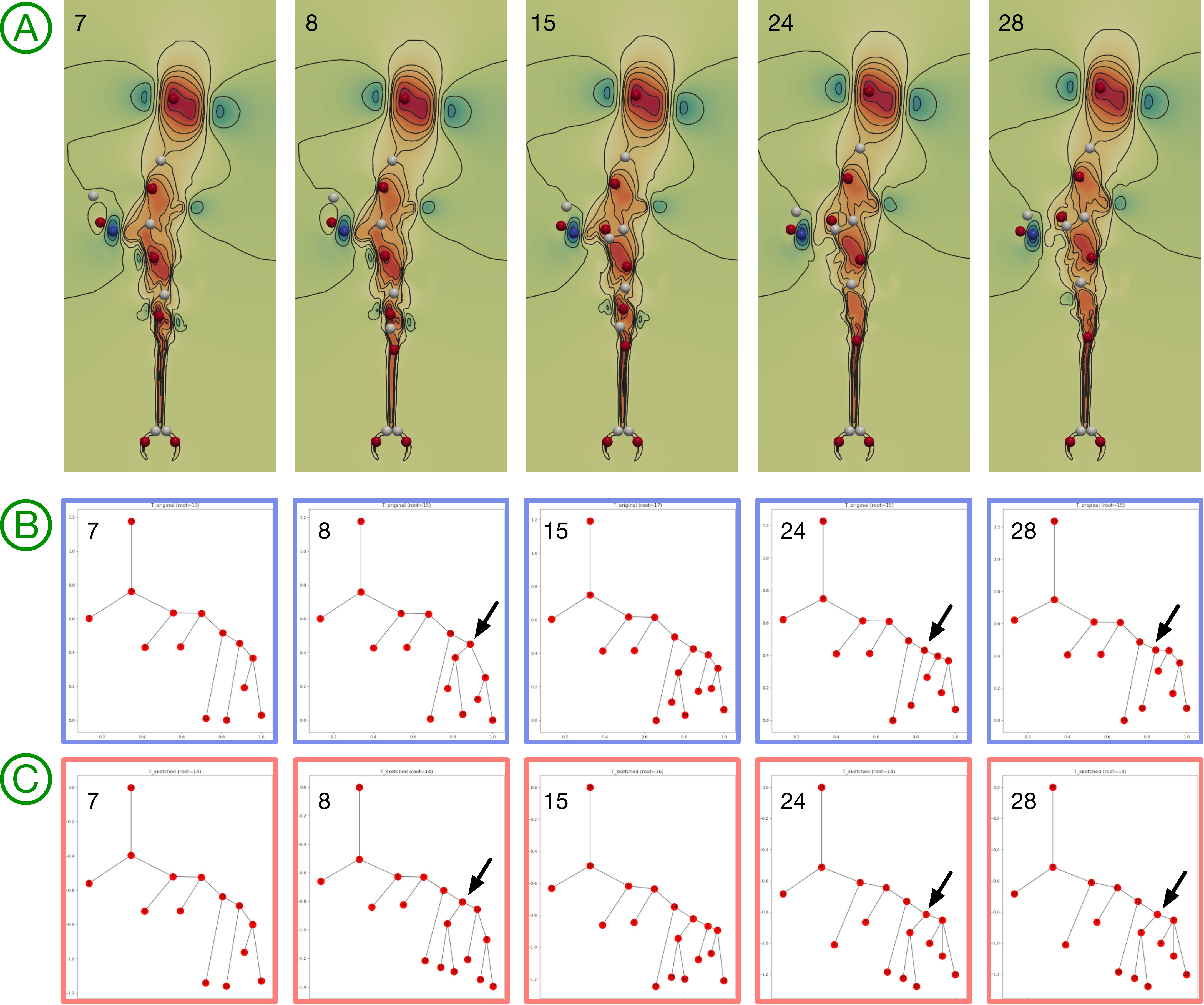}
  \vspace{-2mm}
  \caption{We demonstrate our merge tree sketching framework with a set of 31 merge trees. Each merge tree is generated from the vertical velocity magnitude field coming from a time-varying flow simulation called the \emph{Heated Cylinder}. By combining  matrix sketching -- iterative feature selection -- with probabilistic  matching, we show that the sketched trees (red boxes) approximate the input trees (blue boxes) reasonably well with only three basis trees. (A) A subset of scalar fields that give rise to the input merge trees labeled 7, 8, 15, 24, and 28. (B-C) Comparing  each sketched tree with its corresponding input tree, highlighting  noticeable structural differences among subtrees (whose roots are pointed by black arrows) before and after sketching.}  
  \label{fig:teaser}
}

\vgtcinsertpkg

\begin{document}

\maketitle

\section{Introduction}
\label{sec:intro}

Topological descriptors such as merge trees, contour trees, Reeb graphs, and Morse-Smale complexes serve to describe and identify characteristics associated with scalar fields, with many applications in the analysis and visualization of scientific data (e.g., see the surveys~\cite{HeineLeitteHlawitschka2016, LiuMaljovecWang2017}). 
Sketching, on the other hand, is a class of mathematical tools where a large dataset is replaced by a smaller one that preserves its properties of interests. 
In this paper, we are interested in sketching a set of topological descriptors -- specifically merge trees -- for scientific visualization.
 
We formulate our problem as follows: given a large set $\Tcal$ of merge trees, we would like to find a much smaller basis set $\Scal$ such that each tree in $\Tcal$ can be approximately reconstructed from a linear combination of merge trees in $\Scal$.
The set $\Tcal$ may arise from a time-varying field or an ensemble of scientific simulations, generated with varying parameters and/or different instruments.   
Our motivation is two-fold. We are interested in developing a merge tree sketching framework to:
\begin{itemize} \denselist 
\item Obtain a compressed representation of a large set of merge trees -- as a much smaller set of basis trees together with a coefficient matrix -- for downstream analysis and visualization; 
\item Identify good representatives that capture topological variations in a set of merge trees as well as outliers. 
\end{itemize} 
A sketch of $\Tcal$ with $\Scal$ gives rise to a significantly smaller representation that is a reasonable approximation of $\Tcal$. 
In addition, elements in $\Scal$ will serve as good representatives of $\Tcal$, while elements with large sketching errors will be considered as outliers.

\begin{figure}[!ht]
    \centering
    \includegraphics[width=0.9\columnwidth]{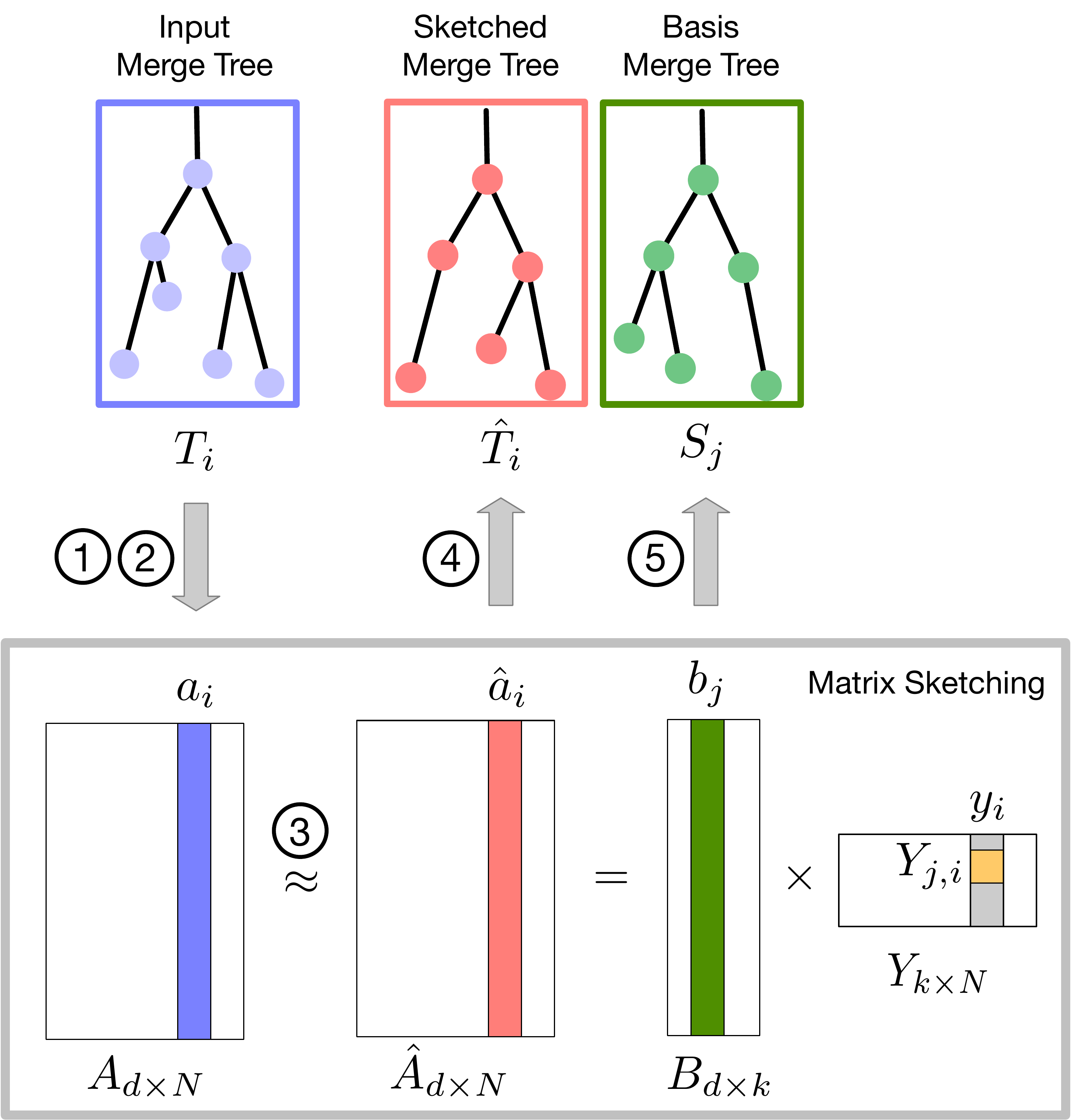}
    \vspace{-2mm}
    \caption{The overall pipeline for sketching a set of merge trees.} 
    \label{fig:pipeline}
\end{figure}

We are inspired by the idea of matrix sketching. A set of high-dimensional vectors is \emph{sketchable} via matrix sketching techniques such as principle component analysis (PCA), and column subset selection (CSS), as illustrated in~\autoref{fig:pipeline} (gray box).  
Given a dataset of $N$ points with $d$ features, represented as a $d \times N$ matrix $A$ (with row-wise zero empirical mean), together with a parameter $k$, PCA aims to find a $k$-dimensional subspace $\Hspace$ of $\Rspace^d$ that minimizes the average squared distance between the points and their corresponding projections onto $\Hspace$.
Equivalently, for every input point $a_i$ (a column vector of $A$), PCA finds a $k$-dimensional embedding $y_i$ (a column vector of $Y$) along the subspace $\Hspace$ to minimize $|| A - \hat{A}||^2_F = ||A-BY||^2_F$.  
$B$ is a $d \times k$ matrix whose columns $b_1, b_2, \dots, b_k$ form an orthonormal basis for $\Hspace$. 
$Y$ is a $k \times N$ coefficient matrix, whose column $y_i$ encodes the coefficients for approximating $a_i$ using the basis from $B$. 
That is, $a_i \approx \hat{a}_i = \sum_{j=1}^{k} b_j Y_{j,i}.$

Another technique we discuss is CSS, whose goal is to find a small subset of the columns in $A$ to form $B$ such that the projection error of $A$ to the span of the chosen columns is minimized, that is, to minimize $|| A - \hat{A}||^2_F = ||A-BY||^2_F$, where we restrict $B$ to come from columns of $A$. 
Such a restriction is important for data summarization, feature selection, and interpretable dimensionality reduction~\cite{BhaskaraLattanziVassilvitskii2019}.  
Thus, with either PCA or CSS, given a set of high-dimensional vectors, we could find a set of basis vectors such that each input vector can be approximately reconstructed from a linear combination of the basis vectors. 

Now, what if we replace a set of high-dimensional vectors by a set of objects that encode topological information of data, specifically topological descriptors?
Up until now, a large set of topological descriptors has not been known to be \emph{sketchable}. 
In this paper, we focus on merge trees, which are a type of topological descriptors that record the connectivity among the sublevel sets of scalar fields.  
We address the following question: given a large set $\Tcal$ of merge trees, can we find a much smaller basis set $\Scal$ as its ``sketch''?

Our overall pipeline is illustrated in~\autoref{fig:pipeline} and detailed in~\autoref{sec:method}.
In steps 1 and 2, given a set of $N$ merge trees $\Tcal=\{T_1,T_2,\cdots, T_N\}$ as input, we represent each merge tree $T_i$ as a metric measure network and employ the Gromov-Wasserstein framework of Chowdhury and Needham~\cite{ChowdhuryNeedham2020} to map it to a column vector $a_i$ in the data matrix $A$. 
In step 3, we apply matrix sketching techniques, in particular, column subset selection (CSS) and non-negative matrix factorization (NMF), to obtain an approximated matrix $\hat{A}$, where $A \approx \hat{A} = B \times Y$.  
In step 4, we convert each column in $\hat{A}$ into a merge tree (referred to as a sketched merge tree) using spanning trees, in particular, minimum spanning trees (MST) or low-stretch spanning trees (LSST).  
Finally, in step 5, we return a set of basis merge trees $\Scal$ by applying LSST or MST to each column $b_j$ in $B$.  Each entry $Y_{j,i}$ in the coefficient matrix $Y$ defines the coefficient for basis tree $S_j$ in approximating $T_i$. 
Thus, intuitively, with the above pipeline, given a set of merge trees, we could find a set of basis trees such that each input tree can be approximately reconstructed from a linear combination of the basis trees. 

Our contribution is two-fold. First, we combine the notion of probabilistic matching via Gromov-Wasserstein distances with matrix sketching techniques to give a class of algorithms for sketching a set of merge trees. 
Second, we provide experimental results that demonstrate the utility of our framework in sketching merge trees that arise from scientific simulations. 
Specifically, we show that understanding the sketchability properties of merge trees can be particularly useful for the study of time-varying fields and  ensembles, where our framework can be used to \emph{obtain compact representations} for downstream analysis and visualizaiton, and to \emph{identify good representatives and outliers}.

\section{Related Work}
\label{sec:related}

We review relevant work on merge trees, Gromov-Wasserstein  distances, graph alignment, matrix sketching, and spanning trees. 

\para{Merge trees}
Merge trees are a type of topological descriptors that record the connectivity among the sublevel sets of scalar fields (see e.g.,~\cite{CarrSnoeyinkAxen2003, BeketayevYeliussizovMorozov2014}). 
They are rooted in Morse theory~\cite{Milnor63}, which characterizes scalar field data by the topological changes in its sublevel sets at isolated critical points. 
In this paper, instead of a direct comparison between a pair of merge trees using existing metrics for merge trees or Reeb graphs (e.g.,~\cite{GasparovicMunchOudot2019,BeketayevYeliussizovMorozov2014, SridharamurthyMasoodKamakshidasan2018}), 
we treat merge trees as metric measure networks and utilize the Gromov-Wasserstein framework described in~\autoref{sec:background} to obtain their alignment and vector representations. 

\para{Gromov-Wasserstein (GW) distances}~Gromov introduced Gromov-Hausdorff (GH) distances~\cite{Gromov1999} while presenting a systematic treatment of metric invariants for Riemannian manifolds.  
GH distances can be employed as a tool for shape matching and comparison (\eg,~\cite{BronsteinBronsteinKimmel2006,Memoli2005,Memoli2007,MemoliSapiro2004,MemoliSapiro2005}), where shapes are treated as metric spaces, and two shapes are considered equal if they are isometric. 
Memoli~\cite{Memoli2011} modified the formulation of GH distances by introducing a relaxed notion of proximity between objects, thus generalizing GH distances to the notion of Gromov-Wasserstein (GW) distances for practical considerations.
Since then, GW distances have had a number of variants based on optimal transport~\cite{TitouanCourtyTavenard2019,TitouanFlamaryCourty2019} and measure-preserving mappings~\cite{MemoliNeedham2020}.  
Apart from theoretical explorations~\cite{Memoli2011,Sturm2012}, GW distances have been utilized in the study of graphs and networks~\cite{Hendrikson2016,XuLuoCarin2019,XuLuoZha2019}, machine learning~\cite{EzuzSolomonKim2017, BunneAlvarez-MelisKrause2019}, and word embeddings~\cite{Alvarez-MelisJaakkola2018}. 
Recently, Memoli \etal~\cite{MemoliSidiropoulosSinghal2018} considered the problem of approximating (sketching) metric spaces using GW distance. 
Their goal was to approximate a (single) metric measure space modeling the underlying data by a smaller metric measure space. 
The work presented in this paper instead focuses on approximating a large set of merge trees -- modeled as a set of  metric measure networks -- with a much smaller set of merge trees.

\para{Aligning and averaging graphs}~Graph alignment or graph matching is a key ingredient in performing comparisons and statistical analysis on the space of graphs  (\eg,~\cite{Emmert-StreibDehmerShi2016,GuMilenkovic2020}).
It is often needed to establish node correspondences between graphs of different sizes. 
The approaches that are most relevant here are the ones based on the GW distances~\cite{PeyreCuturiSolomon2016,ChowdhuryNeedham2020}, which employ \emph{probabilistic matching} (``soft matching'') of nodes. 
Information in a graph can be captured by a symmetric positive semidefinite matrix that encodes distances or similarities between pairs of nodes. 
Dryden \etal~\cite{DrydenKoloydenkoZhou2009} described a way to perform statistical analysis and to compute the mean of such matrices. 
Agueh \etal~\cite{AguehCarlier2011} considered barycenters of several probability measures, whereas Cuturi \etal~\cite{CuturiDoucet2014} and Benamou \etal~\cite{BenamouCarlierCuturi2015} developed efficient algorithms to compute such barycenters. 
Peyre \etal~\cite{PeyreCuturiSolomon2016} combined these ideas with the notion of GW distances~\cite{Memoli2011} to develop GW averaging of distance/similarity matrices.
Chowdhury and Needham~\cite{ChowdhuryNeedham2020} built upon the work in~\cite{PeyreCuturiSolomon2016} and provided a GW framework to compute a Frech\'{e}t mean among these matrices using measure couplings. 
In this paper, we utilize the GW framework~\cite{ChowdhuryNeedham2020} for probabilistic matching  among merge trees. 

\para{Matrix sketching}~Many matrix sketching techniques build upon numerical linear algebra and vector sketching. 
For simplicity, we formulate the problem as follows: 
Given a $d \times N$ matrix $A$, we would like to approximate $A$ using fewer columns, as a $d \times k$ matrix $B$ such that $A$ and $B$ are considered to be \emph{close} with respect to some problem of interest.  
Basic approaches for matrix sketching include truncated SVD, column or row sampling~\cite{DrineasKannanMahoney2006,DrineasMagdon-IsmailMahoney2012}, random projection~\cite{Sarlos2006}, and frequent directions~\cite{GhashamiLibertyPhillips2016, Liberty2013};  see~\cite{Phillips2016, Woodruff2014} for surveys.
 
The column sampling approach carefully chooses a subset of the columns of $A$ proportional to their \emph{importance}, where the importance is determined by the squared norm (\eg,~\cite{DrineasKannanMahoney2006}) or the (approximated) leverage scores (\eg,~\cite{DrineasMagdon-IsmailMahoney2012}). 
The random projection approach takes advantage of the Johnson-Lindenstrauss (JL) Lemma~\cite{JohnsonLindenstrauss1984} to create an $N \times k$ linear projection matrix $S$ (\eg,~\cite{Sarlos2006}), where $B = AS$. 
The frequent directions approach~\cite{Liberty2013,GhashamiLibertyPhillips2016} focuses on replicating properties of the SVD.
The algorithm processes each column of $A$ at a time while maintaining the best rank-$k$ approximation as the sketch.   

\para{Spanning trees of weighted graphs}
Given an undirected, weighted graph $G$, a spanning tree is a subgraph of $G$ that is a tree that connects all the vertices of $G$ with a minimum possible number of edges. 
We consider two types of spanning trees: the \emph{minimal spanning tree} (MST) and the \emph{low stretch spanning tree} (LSST)~\cite{AbrahamBartalNeiman2007,AbrahamBartalNeiman2008,AbrahamNeiman2012}. 
Whereas the MST tries to minimize the sum of edge weights in the tree, LSST tries to minimize the stretch (relative distortion) of pairwise distances between the nodes of $G$.
LSSTs were initially studied in the context of road networks~\cite{AlonKarpPeleg1995}. They also play an important role in fast solvers for symmetric diagonally dominant (SDD) linear systems~\cite{ElkinEmekSpielman2005,KoutisMillerPeng2011}.

\section{Technical Background}
\label{sec:background}

We begin by reviewing the notion of a merge tree that arises from a scalar field. 
We then introduce the technical background needed to map a merge tree to a column vector in the data matrix. 
Our framework utilizes the probabilistic matching from the Gromov-Wasserstein (GW) framework of Chowdhury and Needham~\cite{ChowdhuryNeedham2020}, with a few ingredients from Peyre \etal~\cite{PeyreCuturiSolomon2016}. 

\begin{figure}[!ht]
    \centering
    \includegraphics[width=0.98\columnwidth]{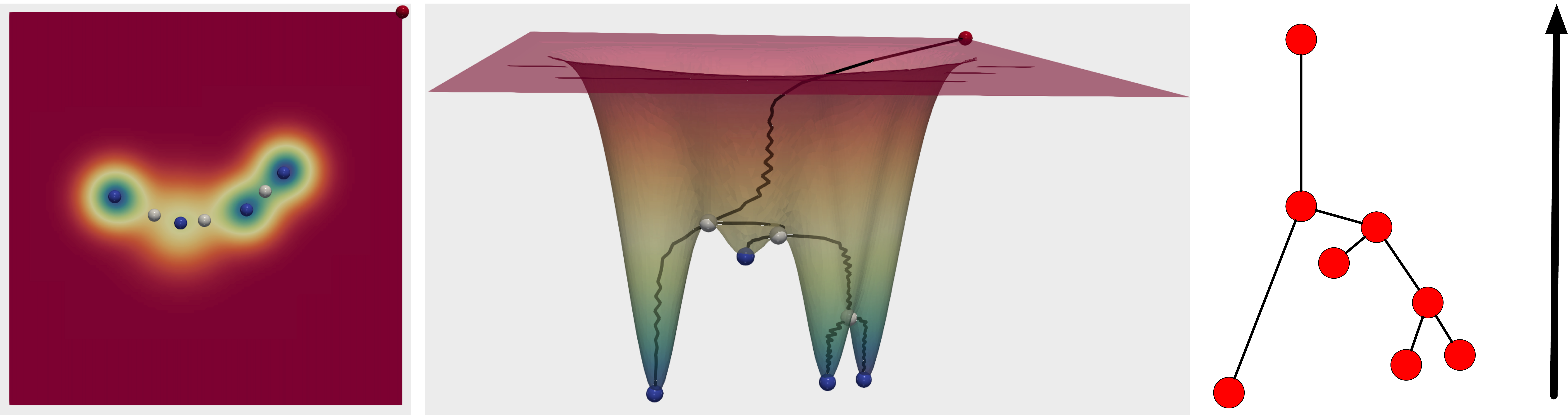}
    \vspace{-2mm}
    \caption{An examples of a merge tree from a height field. From left to right: 2D scalar fields visualization, merge trees embedded in the graphs of the scalar fields, and abstract visualization of merge trees as rooted trees equipped with height functions.} 
    \label{fig:mt}
    \vspace{-4mm}
\end{figure}

\para{Merge trees}~Let $f: \Mspace \to \Rspace$ be a scalar field defined on the domain of interest $\Mspace$, where $\Mspace$ can be a manifold or a subset of $\Rspace^d$.
For our experiments in~\autoref{sec:results}, $\Mspace \subset \Rspace^2$. 
Merge trees capture the connectivity among the \emph{sublevel sets} of $f$, \ie, $\Mspace_a = f^{-1}(-\infty, a]$. 
Formally, two points $x, y \in \Mspace$ are \emph{equivalent}, denoted by $x \sim y$, if $f(x) = f(y)=a$, and $x$ and $y$ belong to the same connected component of a sublevel set $\Mspace_a$. 
The \emph{merge tree}, $T(\Mspace, f) = \Mspace/{\sim}$, is the quotient space obtained by gluing together points in $\Mspace$ that are equivalent under the relation $\sim$.
To describe a merge tree procedurally, as we sweep the function value $a$ from $-\infty$ to $\infty$, we create a new branch originating at a leaf node for each local minimum of $f$. 
As $a$ increases, such a branch is extended as its corresponding component in $\Mspace_a$ grows until it merges with another branch at a saddle point. 
If $\Mspace$ is connected, all branches eventually merge into a single component at the global maximum of $f$, which corresponds to the root of the tree. 
For a given merge tree, leaves, internal nodes, and root node represent the minima, merging saddles, and global maximum of $f$, respectively.  
\autoref{fig:mt} displays a scalar field with its corresponding merge tree embedded in the graph of the scalar field. 
Abstractly, a merge tree $T$ is a rooted tree equipped with a scalar function defined on its node set,  $f: V \to \Rspace$. 

\para{Gromov-Wasserstein distance for measure networks} The GW  distance was proposed by Memoli~\cite{Memoli2007,Memoli2011} for metric measure spaces. 
Peyre \etal~\cite{PeyreCuturiSolomon2016} introduced the notion of a \emph{measure network} and defined the GW distance between such networks. 
The key idea is to find a \emph{probabilistic matching} between a pair of networks by searching over the convex set of couplings of the  probability measures defined on the networks.

A finite, weighted graph $G$ can be represented as a measure network using a triple $(V, W, p)$, where $V$ is the set of $n$ nodes, $W$ is a weighted adjacency matrix, and $p$ is a probability measure supported on the nodes of $G$.
For our experiments, $p$ is taken to be uniform, that is, $p = \frac{1}{n} \onevec{n}$, where $\onevec{n} = (1,1,\dots, 1)^{T} \in \Rspace^{n}$.

Let $G_{1}(V_1, W_{1},p_{1})$ and $G_{2}(V_2, W_{2},p_{2})$ be a pair of graphs with $n_{1}$ and $n_2$ nodes, respectively. 
Let $[n]$ denote the set $\{1,2,\dots,n\}$.
$V_1=\{x_i\}_{i \in [n_1]}$ and $V_2=\{y_j\}_{j \in [n_2]}$. 
A \emph{coupling} between probability measures $p_{1}$ and $p_{2}$ is a joint probability measure on $V_1 \times V_2$ whose marginals agree with $p_1$ and $p_2$. 
That is, a coupling is represented as an $n_{1}\times n_{2}$ non-negative matrix $C$ such that $C\onevec{n_{2}} = p_{1}$ and $C^{T}\onevec{n_{1}} = p_{2}$. 
Given matrix $C$, its \emph{binarization} is an $n_{1}\times n_{2}$ binary matrix, denoted as $\onevec{C > 0}$: this matrix has $1$ where $C>0$, and $0$ elsewhere. 

The \emph{distortion} of a coupling $C$ with an arbitrary loss function $L$ is defined as~\cite{PeyreCuturiSolomon2016} 
\begin{align}
\distortion(C) = \sum_{ i,k \in [n_1],  j,l \in [n_2]} L(W_1(i,k), W_{2}(j,l))C_{i,j}C_{k,l}.
\label{eq:distortion}
\end{align} 

Let $\Ccal = \Ccal(p_1, p_2)$ denote the collection of all couplings between $p_{1}$ and $p_{2}$.
The \emph{Gromov-Wasserstein discrepancy}~\cite{PeyreCuturiSolomon2016} is defined as 
\begin{align}
\discrepancy(C) = \min_{C\in\Ccal} \distortion(C).
\label{eq:discrepancy}
\end{align} 

In this paper, we consider the quadratic loss function $L(a,b) = \frac{1}{2}|a-b|^{2}$. 
The \emph{Gromov-Wasserstein distance}~\cite{ChowdhuryNeedham2020, Memoli2011, PeyreCuturiSolomon2016} $d_{GW}$ between $G_{1}$ and $G_{2}$ is defined as
\begin{align}
d_{GW}(G_{1},G_{2}) = \frac{1}{2} \min_{C\in\Ccal} \sum_{ i,k \in [n_1],  j,l \in [n_2]} |W_1(i,k) - W_{2}(j,l)|^2C_{i,j}C_{k,l}.
\label{eq:gw-distance}
\end{align}
It follows from the work of Sturm~\cite{Sturm2012} that such minimizers always exist and are referred to as \emph{optimal couplings}. 

\para{Alignment and blowup}~Given a pair of graphs $G_{1} = (V_1, W_1, p_1)$ and $G_{2} = (V_2, W_2, p_2)$ with $n_1$ and $n_2$ nodes respectively, a coupling $C \in \Ccal(p_{1},p_{2})$ can be used to \emph{align} their nodes.
In order to do this, we will need to increase the size of $G_1$ and $G_2$ appropriately into their respective \emph{blowup} graphs $G'_1$ and $G'_2$, such that $G'_1$ and $G'_2$ contain the same $n$ number of nodes (where $n_1, n_2 \le n$).  
Roughly speaking, let $x$ be a node in $G_1$, and let $n_x$ be the number of nodes in $G_2$ that have a nonzero coupling probability with $x$. 
The blowup graph $G'_1 = (V'_1, W'_1, p'_1)$ is created by making $n_x$ copies of node $x$ for each node in $G_1$, generating a new node set $V'_1$. The probability distribution $p'_1$ and the weight matrix $W'_1$ are updated from $p_1$ and $W_1$ accordingly. 
 Similarly, we can construct the blowup $G'_2 = (V'_2, W'_2, p'_2)$ of $G_{2}$. 

An optimal coupling $C$ expands naturally to a coupling $C'$ between $p'_1$ and $p'_2$. 
After taking appropriate blowups, $C'$ can be binarized to be an $n \times n$ permutation matrix, and used to align the nodes of the two blown-up graphs. 
The GW distance is given by a formulation equivalent to~\autoref{eq:gw-distance} based on an optimal coupling, 
\begin{align}
d_{GW}(G_{1}, G_{2}) = \frac{1}{2} \sum_{i,j} |W'_1(i,j) - W'_{2}(i,j)|^2 p'_1(i)p'_1(j).
\label{eq:gw-distance-blowup}
\end{align}

\para{Fr\'{e}chet mean}~Given a collection of graphs $\Gcal = \{G_1, G_2, \dots, G_N\}$,  a \emph{Fr\'{e}chet mean}~\cite{ChowdhuryNeedham2020} $\overline{G}$ of $\Gcal$ is a minimizer of the functional $F(H, \Gcal) = \frac{1}{N} \sum_{i=1}^{N} d_{GW}(G_{i},H)$ over the space $\Ncal$ of measure networks, 
\begin{align}
\overline{G} = \min_{H \in \Ncal} \frac{1}{N} \sum_{i=1}^{N} d_{GW}(G_{i},H).
\label{eq:mean}
\end{align}

Chowdhury and Needham~\cite{ChowdhuryNeedham2020} defined the directional derivative and the gradient of the functional $F(H, \Gcal)$ at $H$ and provided a gradient descent algorithm to compute the Fr\'{e}chet mean.
Their iterative optimization begins with an initial guess $H_0$ of the Fr\'{e}chet mean.
At the $k^{th}$ iteration, there is a two-step process: each $G_i$ is first blown-up and aligned to the current Fr\'{e}chet mean, $H_k$; then $H_k$ is updated using the gradient of the functional $F(H_k, \Gcal)$ at $H_k$. 
Such a two-step process is repeated until convergence where the gradient vanishes.
For the complete algorithmic and implementational details, see~\cite{ChowdhuryNeedham2020}.
If $\overline{G}=(\overline{V}, \overline{W}, \overline{p})$ is the Fr\'{e}chet mean, then we have
\[\overline{W}(i,j) = \frac{1}{N} \sum_{k=1}^{N} W'_{k}(i,j) ,\]
where $W'_k$ is the weight matrix obtained by blowing-up and aligning $G_k \in \Gcal$ to $\overline{G}$.
That is, when all the graphs in $\Gcal$ are blown-up and aligned to $\overline{G}$, the weight matrix of $\overline{G}$ is given by a simple element-wise average of the weight matrices of the graphs. 

\begin{figure}[!ht]
    \centering
    \includegraphics[width=0.98\columnwidth]{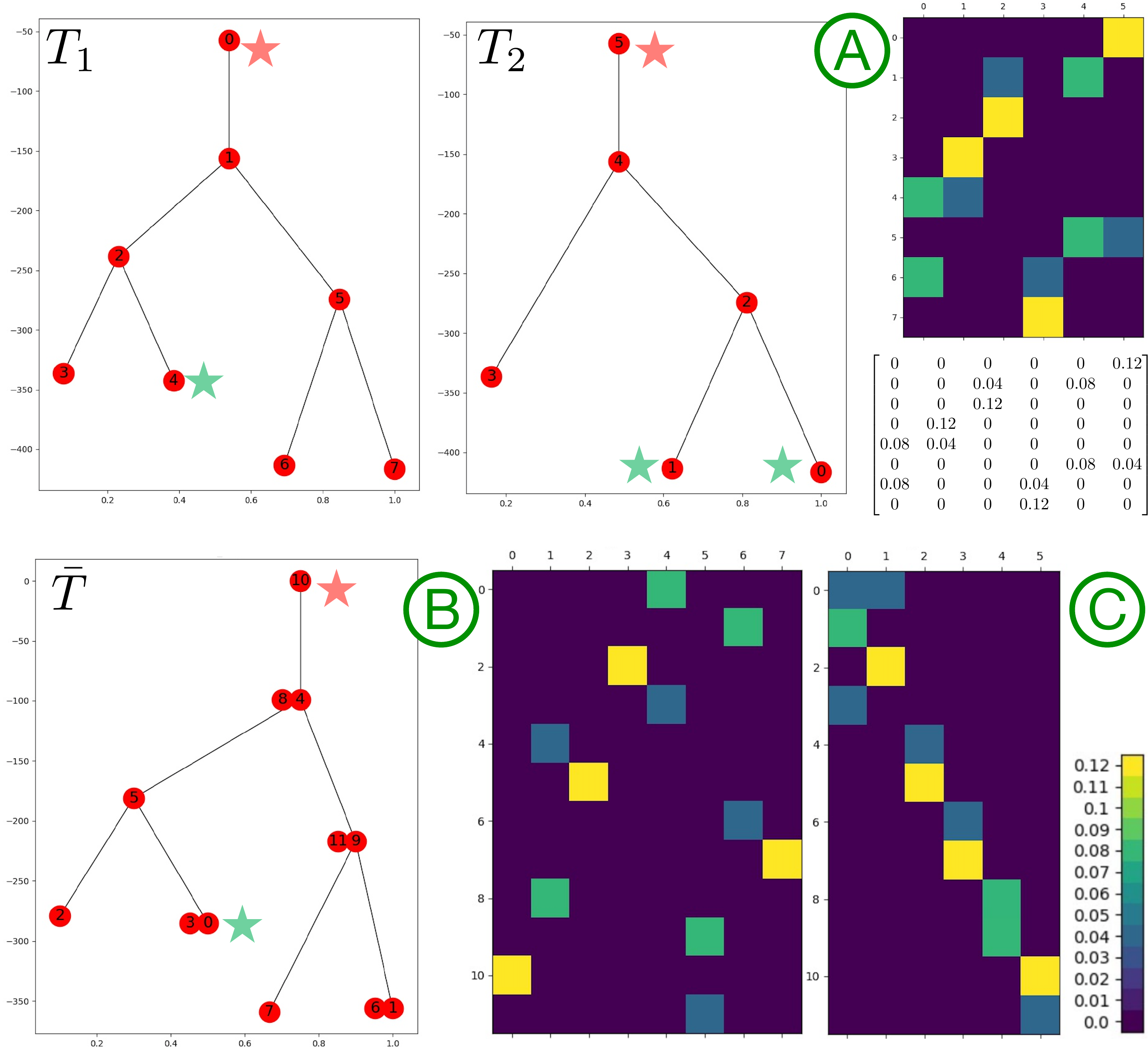}
    \vspace{-2mm}
    \caption{An optimal coupling between two simple merge trees $T_1$ and $T_2$. The coupling matrix is visualized in (A): yellows means high and dark blue means low probability. Couplings between the Fr\'{e}chet mean $\overline{T}$ with $T_1$ and $T_2$ are shown in (B) and (C), respectively. } 
    \label{fig:coupling}
\end{figure}

\para{A simple example}~We give a simple example involving a pair of merge trees in~\autoref{fig:coupling}.
$T_1$ and $T_2$ contain 8 and 6 nodes, respectively (nodes are labeled starting with a $0$ index). 
The optimal coupling $C$ obtained by the gradient descent algorithm  is visualized in \autoref{fig:coupling}(A).  
$C$ is an $8 \times 6$ matrix, and it shows that node 0 in $T_1$ is matched to node 5 in $T_2$ with the highest probability ($0.12$, red stars). 
Node 4 in $T_1$ is coupled with both node 0 (with a probability $0.08$) and node 1 (with a probability $0.04$)  in $T_2$ (green stars).

Now, we compute the Fr\'{e}chet mean $\overline{T}$ of $T_1$ and $T_2$, which has 12 nodes. 
We align both $T_1$ and $T_2$ to $\overline{T}$ via their blowup graphs. 
This gives rise to a coupling matrix between $\overline{T}$ and $T_1$ (of size $12 \times 8$) in \autoref{fig:coupling}(B), and a coupling matrix between $\overline{T}$ and $T_2$ (of size $12 \times 6$) in \autoref{fig:coupling}(C), respectively.   
As shown in~\autoref{fig:coupling}, root node 10 of $\overline{T}$ is matched with root node 0 of $T_1$ and root node 5 of $T_2$ (red stars). 
Node 0 of $\overline{T}$ is matched probabilistically with node 4 in $T_1$ and nodes $0$ and $1$ in $T_2$ (green stars). 
Now both trees $T_1$ and $T_2$ are blown-up to be $T'_1$ and $T'_2$, each with 12 nodes, and can be vectorized into column vectors of the same size.

\section{Methods}
\label{sec:method}
Given a set $\Tcal$ of $N$ merge trees as input, our goal is to find a basis set $\Scal$ with $k \ll N$ merge trees such that each tree in $\Tcal$ can be approximately reconstructed from a linear combination of merge trees in $\Scal$.
We propose to combine the GW framework~\cite{ChowdhuryNeedham2020} with techniques from matrix sketching to achieve this goal.
We detail our pipeline to compute $\Scal$, as illustrated in~\autoref{fig:pipeline}. 

\begin{figure*}[h]
    \centering{
    \includegraphics[width=1.7\columnwidth]{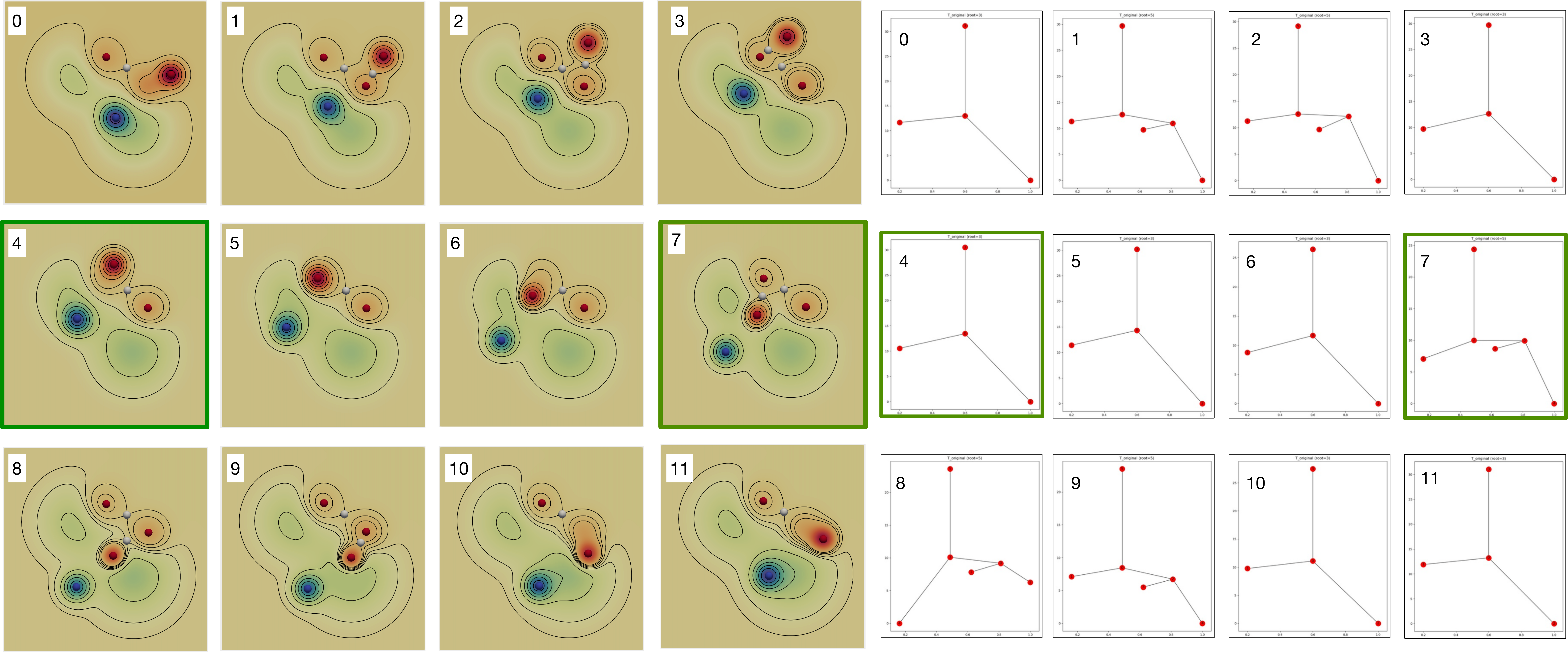}
    \vspace{-2mm}
    \caption{Visualizing a time-varying mixture of Gaussian functions (left) together with (right) their corresponding merge trees.} 
    \label{fig:Gaussian-input}}
\end{figure*}

\para{Step~1:~Representing merge trees as measure networks}
The first step is to represent merge trees as metric measure networks as described in~\autoref{sec:background}.
Each merge tree $T \in \Tcal$ can be represented using a triple $(V, W, p)$, where $V$ is the node set, $W$ is a matrix of pairwise distances between its nodes, and $p$ is a probability distribution on $V$.  

In this paper, we define $p$ as a uniform distribution, \ie, $p = \frac{1}{|V|} \onevec{|V|}$. 
Recall that each node $x$ in a merge tree is associated with a scalar value $f(x)$.
For a pair of nodes $x, x' \in V$, if they are adjacent, we define $W(x, x')=|f(x) - f(x')|$, {\ie}, their absolute difference in function value; otherwise, $W(x, x')$ is the shortest path distance between them in $T$. 
By construction, a shortest path between two nodes goes through their lowest common ancestor in $T$. 
We define $W$ in such a way to encode information in $f$, which is inherent to a merge tree.

\para{Step~2:~Merge tree vectorization via alignment to the Fr{\'e}chet mean} 
The second step is to convert each merge tree into a column vector of the same size via blowup and alignment to the Fr{\'e}chet mean. 
Having represented each merge tree as a metric measure network, we can use the GW framework to compute a Fr{\'e}chet mean of $\Tcal$, denoted as $\overline{T} = (\overline{V}, \overline{W}, \overline{p})$.
Let $n = |\overline{V}|$. 
In theory, $n$ may become as large as $\prod_{i=1}^{N}|V_i|$. 
In practice, $n$ is chosen to be much smaller; in our experiment, we choose $n$ to be a small constant factor (2 or 3) times the size of the largest input tree. 
The optimal coupling $C$ between $\overline{T}$ and $T_i$ is an $n \times n_i$ matrix with at least $n$ nonzero entries. 
If the number of nonzero entries in each row is greater than $n$, we keep only the largest value.
That is, if a node of $\overline{T}$ has a nonzero probability of coupling with more than one node of $T$, we consider the mapping with only the highest  probability, so that each coupling matrix $C$ has exactly $n$ nonzero entries.  
We then blow up each $T$ to obtain $T' = (V', W', p')$, and align $\overline{T}$ with $T'$. 
The above procedure ensures that each blown-up tree $T'$ has exactly $n$ nodes, and the binarized coupling matrix $C'$ between $\overline{T}$ and  $T'$ induces a node matching between them. 

We can now vectorize (\ie,~flatten) each $W'$ (an $n\times n$ matrix) to form a column vector $a \in \Rspace^{d}$ of matrix $A$ (where $d = n^2$), as illustrated in~\autoref{fig:pipeline} (step 2)\footnote{In practice, $d = (n+1)n/2$ as we store only the upper triangular matrix.}.
Each $a$ is a vector representation of the input tree $T$ with respect to the Fr{\'e}chet mean $\overline{T}$.

\para{Step 3: Merge tree sketching via matrix sketching}
The third step is to sketch merge trees by applying matrix sketching to the data matrix $A$, as illustrated in~\autoref{fig:pipeline} (step 3). 
By construction, $A$ is a $d \times N$ matrix whose column vectors $a_i$ are vector representations of $T_i$. 
We apply matrix sketching techniques to approximate $A$ by $\hat{A} = B \times Y$. 
In our experiments, we use two linear sketching techniques, namely, column subset selection (CSS) and non-negative matrix factorization (NMF). See~\autoref{sec:implementation} for implementation details.
 
Using CSS, the basis set is formed by sampling $k$ columns of $A$.
Let $B$ denote the matrix formed by $k$ columns of $A$ and let $\Pi = B B^{\pinv}$ denote the projection onto the $k$-dimensional space spanned by the columns of $B$.
The goal of CSS is to find $B$ such that $\lVert A- \Pi A \rVert_F$ is minimized.
We experiment with two variants of CSS.

In the first variant of CSS, referred to as \emph{Length Squared Sampling} ({\LSS}), we sample (without replacement) columns of $A$ with probabilities $q_i$ proportional to the square of their Euclidean norms, \ie, $q_i = {\lVert a_i\rVert_2^2}/{\lVert A\rVert^2_F}$.
We modify the algorithm slightly such that before selecting a new column, we factor out the effects from columns that are already chosen, making the chosen basis as orthogonal as possible.  

In the second variant of CSS, referred to as the \emph{Iterative Feature Selection} ({\IFS}), we use the algorithm proposed by Ordozgoiti \etal~\cite{OrdozgoitiCanavalMozo2016}.
Instead of selecting columns sequentially as in {\LSS}, {\IFS} starts with a random subset of $k$ columns. Then each selected column is either kept or replaced with another column, based on the residual after the other selected columns are factored out simultaneously. 

In the case of NMF, the goal is to compute non-negative matrices $B$ and $Y$ such that $\lVert A - \hat{A}\rVert_F = \lVert A - BY \rVert_F$ is minimized.
We use the implementation provided in the decomposition module of the \textsf{scikit-learn} package~\cite{CichockiPhan2009, FevotteIdier2011}.
The algorithm initializes matrices $B$ and $X = Y^T$ and minimizes the residual $Q = A - BX^T + b_jx_j^T$ alternately with respect to column vectors $b_j$ and $x_j$ of $B$ and $X$, respectively, subject to the constraints $b_j \ge 0$ and $x_j \ge 0$. 

\para{Step 4: Reconstructing sketched merge trees}
For the fourth step, we convert each column in $\hat{A}$ as a sketched merge tree.    
Let $\hat{A} = BY$, where matrices $B$ and $Y$ are obtained using CSS or NMF.
Let $\hat{a} = \hat{a}_i$ denote the $i^{th}$ column of $\hat{A}$.
We reshape $\hat{a}$ as an $n \times n$ weight matrix $\hat{W}'$. 
We then obtain a tree structure $\hat{T}'$ from $\hat{W}'$ by computing its MST or LSST. 

A practical consideration is the \emph{simplification} of a sketched tree $\hat{T}'$  coming from NMF. 
$\hat{T}'$ without simplification is an approximation of the blow-up tree $T'$.
It contains many more nodes compared to the original tree $T$.
Some of these are internal nodes with exactly one parent node and one child node.
In some cases, the distance between two nodes is almost zero.
We further simplify $\hat{T}'$ to obtain a final sketched tree $\hat{T}$ by removing internal nodes and nodes that are too close to each other; see~\autoref{sec:implementation} for details.

\para{Step 5: Returning basis trees}
Finally, we return a set of basis merge trees $\Scal$ using information encoded in the matrix $B$.
Using CSS, each column $b_j$ of $B$ corresponds directly to a column in $A$; therefore, the set $\Scal$ is trivially formed by the corresponding merge trees from $\Tcal$. 
Using NMF, we obtain each basis tree by applying MST or LSST to columns $b_j$ of $B$ with appropriate simplification, as illustrated in~\autoref{fig:pipeline} (step 5).

\para{Error analysis}
For each of our experiments, we compute the \emph{global sketch error} $\epsilon = \lVert A  - \hat{A} \rVert^2_F$, as well as \emph{column-wise sketch error} $\epsilon_i = \lVert a_i - \hat{a}_i\rVert^2_2$, where $\epsilon = \sum_{i=1}^{N} \epsilon_i$.  
By construction, $\epsilon_i \leq \epsilon$. 
For merge trees,  we measure the GW distance between each tree $T_i$ and its sketched version $\hat{T}_i$, that is $\tau_i = d_{GW}(T_i, \hat{T}_i)$, referred to as the \emph{column-wise  GW loss}. 
The \emph{global GW loss} is defined to be $\tau = \sum_{i=1}^{N} \tau_i$.  
For theoretical considerations, see discussions in~\autoref{sec:theory}.

\para{A simple synthetic example}
We give a simple synthetic example to illustrate our pipeline. 
A time-varying scalar field $f$ is a mixture of 2D Gaussians that translate and rotate on the plane. 
We obtain a set $\Tcal=\{T_0, \dots, T_{11}\}$ of merge trees from 12 consecutive time steps, referred to as the \emph{Rotating Gaussian} dataset. 
In~\autoref{fig:Gaussian-input}, we show the scalar fields and the corresponding merge trees, respectively. 
Each merge tree is computed from $-f$; thus, its leaves correspond to the local maxima (red), internal nodes are saddles (white), and the root node is the global minimum (blue) of $f$.

\begin{figure}[!ht]
    \centering
    \includegraphics[width=0.98\columnwidth]{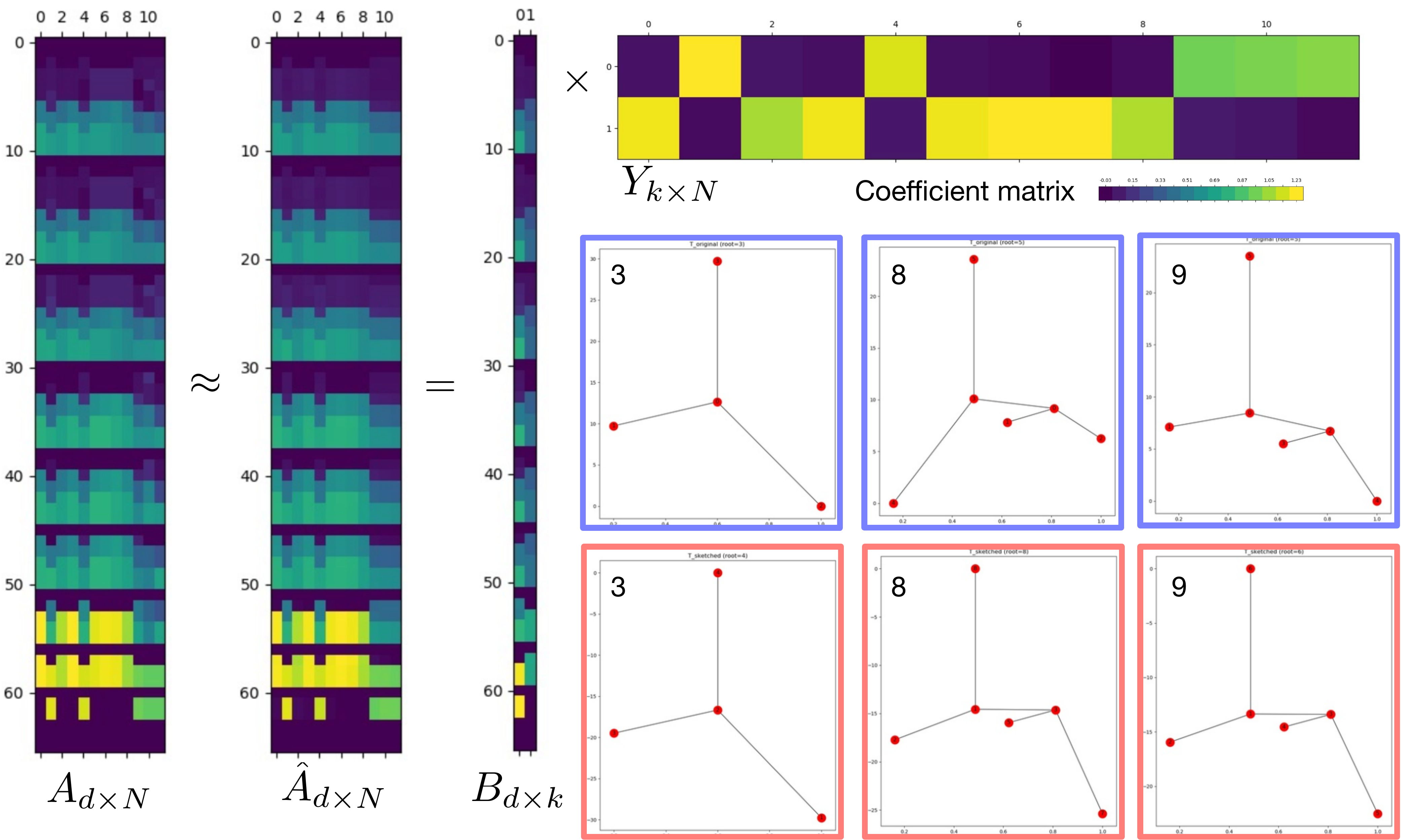}
    \caption{\emph{Rotating Gaussian} dataset: Examples of input merge trees (blue boxes) with their sketched versions (red boxes). Visualizing data matrices associated with the sketching, while highlighting the coefficient matrix.} 
    \label{fig:Gaussian-IFS}
\end{figure}

Since the dataset is quite simple, a couple of basis trees are sufficient to obtain very good sketching results. 
Using $k=2$, {\IFS} select $\Scal = \{T_4, T_7\}$.
In~\autoref{fig:Gaussian-input}, we highlight the two basis trees selected with {\IFS} and their corresponding scalar fields, respectively, with green boxes. 
The topological structures of $T_1$ and $T_6$ are noticeably distinct among the input trees.  
They clearly capture the structural variations and serve as good representatives of the set $\Tcal$.  

We also show a few input trees $T_3, T_8, T_9$ (blue boxes) and their sketched versions (red boxes) in~\autoref{fig:Gaussian-IFS}.  
The input and the sketched tree for $T_3$ are almost indistinguishable. However, there are some structural differences between the input and sketched trees for $T_8$ and $T_9$ due to  randomized approximations. 
We also visualize the data matrix $A$, $\hat{A}$, $B$, and highlight the coefficient matrix $Y$ in~\autoref{fig:Gaussian-IFS}. 
The Frech\'{e}t mean tree $\overline{T}$ contains $11$ nodes. 
The coefficient matrix, shows that each input tree (column) is well represented (with high coefficient) by one of the two basis trees. 
In particular, columns in the coefficience matrix with high (yellow or light green) coefficients ({\wrt} the given basis) may be grouped together, forming two clusters $\{T_0, T_2, T_3, T_5, T_6, T_7. T_8\}$ and $\{T_1, T_4, T_9, T_{10}, T_{11}\}$ whose elements look structurally similar.

 \begin{figure}[!ht]
    \centering
    \includegraphics[width=0.99\columnwidth]{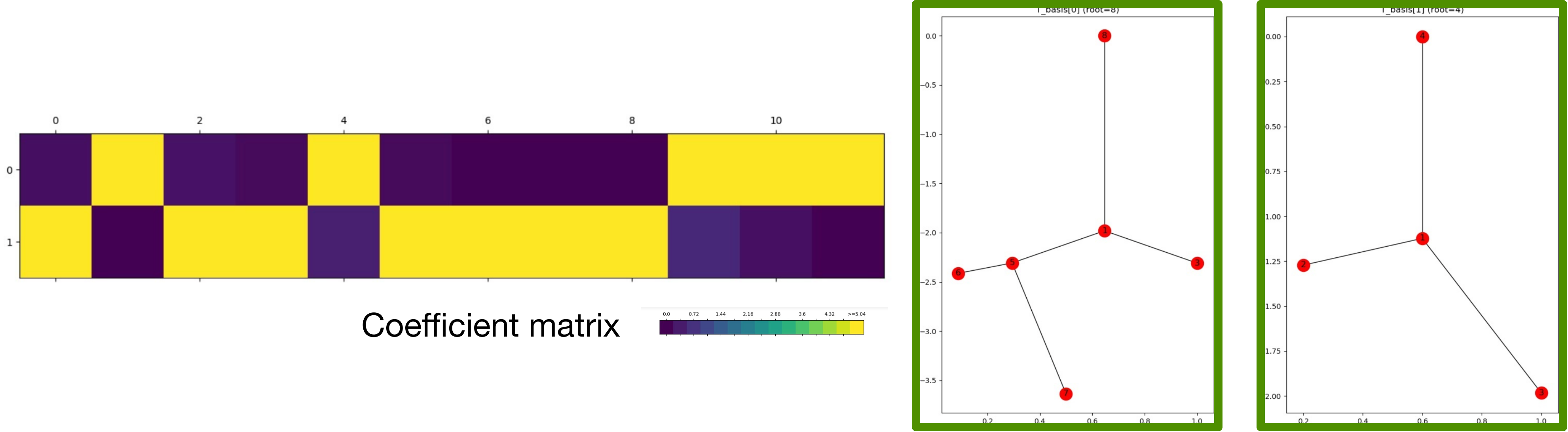}
    \caption{\emph{Rotating Gaussian} data set: coefficient matrices  together with basis trees returned by NMF. } 
    \label{fig:Gaussian-NMF}
\end{figure}

On the other hand, using NMF, when $k=2$, we display the coefficient matrix together with basis trees (obtained via MST) in~\autoref{fig:Gaussian-NMF}. 
The most interesting aspect of using NMF is that the basis trees (green boxes) are not elements of $\Tcal$; however, they very much resemble the basis trees obtained by {\IFS}. 
In addition, columns in the coefficient matrix with high coefficients ({\wrt} the same basis) may be grouped together that show the same two clusters as before.

\section{Experimental Results}
\label{sec:results}

We demonstrate the applications of our sketching framework with  merge trees that arise from three time-varying datasets from scientific simulations. 
The key takeaway is that, using matrix sketching and probabilistic   matching between the merge trees, a large set $\Tcal$ of merge trees is replaced by a much smaller basis set $\Scal$ such that trees in $\Tcal$ are well approximated by trees in $\Scal$. 
Such a compressed representation can then be used for downstream analysis and visualization. 
In addition, our framework makes each large dataset simple to understand, where elements in $\Scal$ serve as good representatives that capture structural variations among the time instances, and elements with large sketching errors are considered as outliers ({\wrt} a chosen basis). 

\begin{figure}[!ht]
    \centering
    \includegraphics[width=0.48\textwidth]{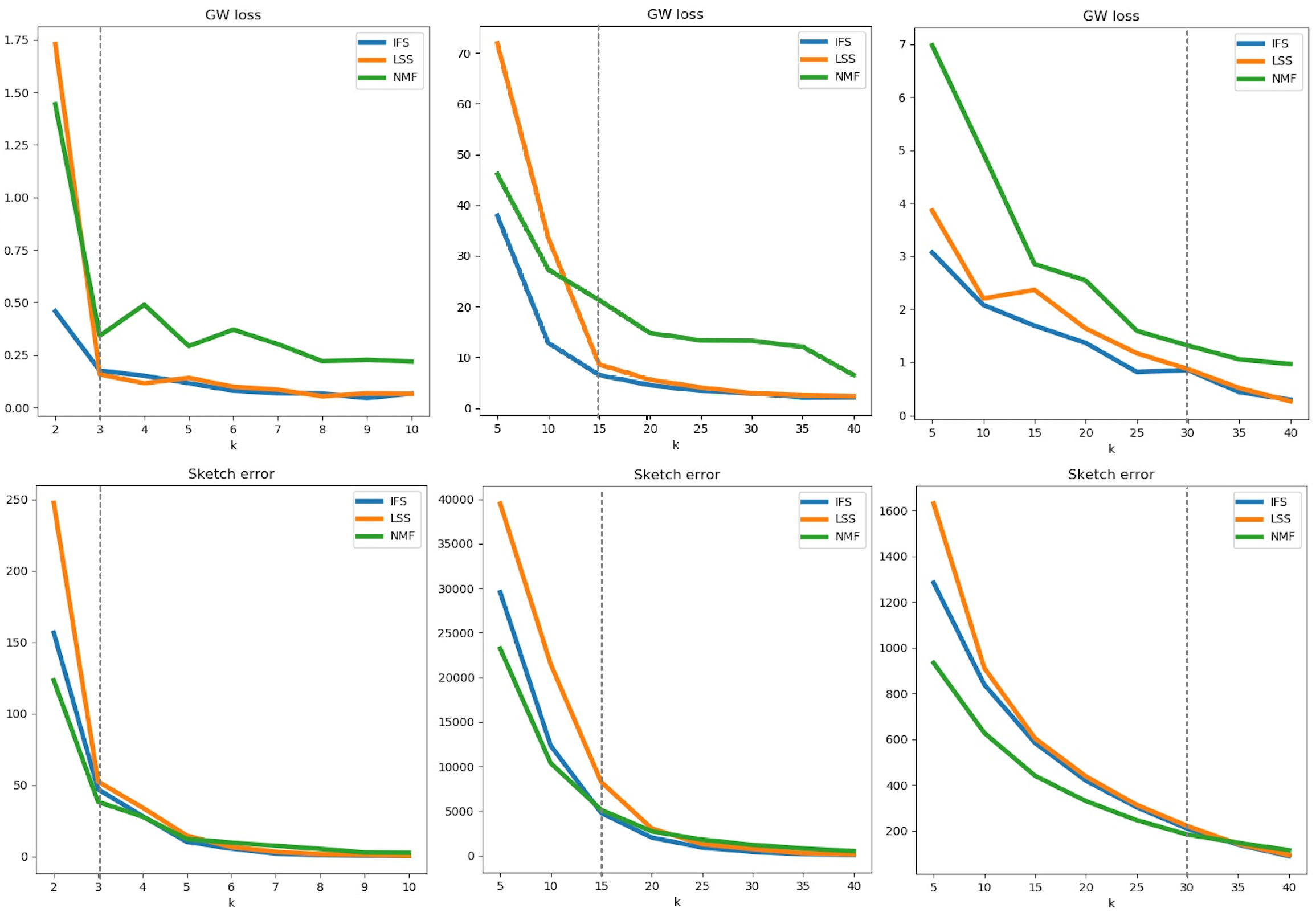}
    \caption{Global GW losses and global sketch errors for varying $k$, the number of basis trees. From left to right, \emph{Heated Cylinder}, \emph{Corner Flow}, and \emph{Red Sea} datasets.} 
    \label{fig:elbow}
\end{figure}

\para{Parameters}
To choose the appropriate $k$ number of basis trees for each dataset, we use the ``elbow method'' to determine $k$, similar to cluster analysis. 
We plot the global GW loss and global sketch error as a function of $k$, and pick the elbow of the curve as the $k$ to use. 
As shown in~\autoref{fig:elbow}, $k$ is chosen to be 3, 15 and 30 for the \emph{Heated Cylinder}, \emph{Corner Flow}, and \emph{Red Sea}\footnote{Admittedly, \emph{Red Sea} dataset is the hardest to sketch, even 30 (basis trees) may not be the optimal.} datasets respectively. 
In subsequent sections, element-wise GW losses and sketch errors also reaffirm these choices. 

For our experiments shown in~\autoref{fig:elbow},  {\IFS} and {\LSS} give  sketched trees with lower GW losses than NMF, in particular, for datasets with smaller merge trees or smaller amount of topological changes, such as \emph{Heated Cylinder} and \emph{Corner Flow} datasets.
While NMF performs better for \emph{Red Sea} dataset with lower sketch errors when individual input trees do not capture the complex topological changes across time instances. 
Furthermore, {\IFS} (blue curve) performs slightly better than {\LSS} (orange curve), based on error analysis (see~\autoref{fig:elbow} and~\autoref{sec:error-analysis} for details).
In term of merge tree reconstruction from distance matrices, MST  generally gives more visually appealing sketched trees and basis trees in practice than LSST, thus we discuss MSTs throughout this section and include some results on LSST in~\autoref{sec:lsst}.

\subsection{Heated Cylinder Dataset}

Two of our datasets come from numerical simulations available online\footnote{https://cgl.ethz.ch/research/visualization/data.php}. 
The first dataset, referred to as the \emph{Heated Cylinder with Boussinesq Approximation} (\emph{Heated Cylinder} in short), comes from the simulation of a 2D flow generated by a heated cylinder using the Boussinesq approximation~\cite{GuntherGrossTheisel2017, gerrisflowsolver}. 
The dataset shows a time-varying turbulent plume containing numerous small vortices.  
We convert each time instance of the flow (a vector field) into a scalar field using the magnitude of its vertical (y) velocity component. 
We generate a set of merge trees from these scalar fields based on 31 time steps -- they correspond to steps 600-630 from the original 2000 time steps. 
This set captures the evolution of small vortices over time. 

\begin{figure}[!ht]
    \centering
    \includegraphics[width=1.0\columnwidth]{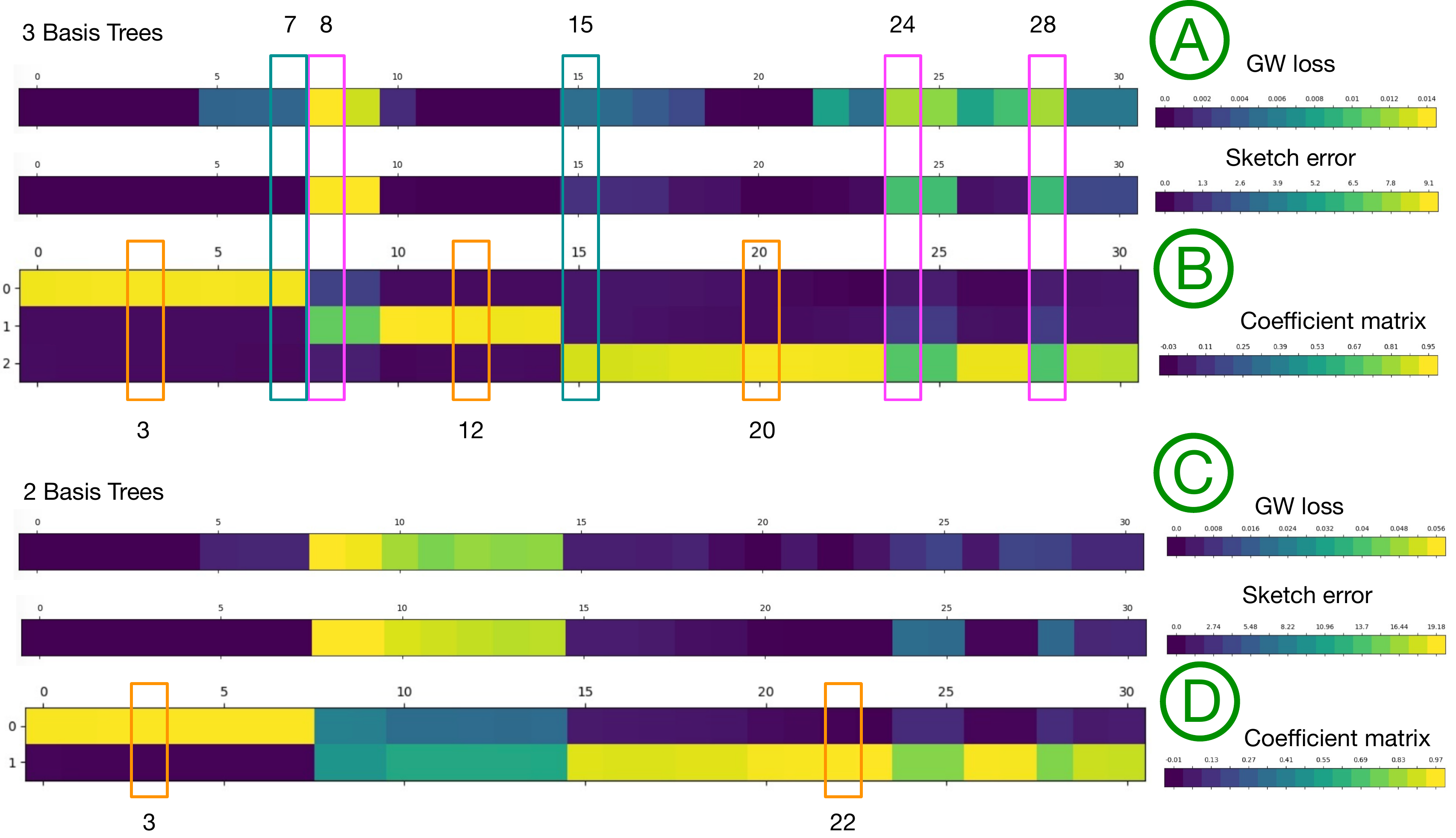}
    \vspace{-6mm}
    \caption{Sketching the \emph{Heated Cylinder} dataset with three (A-B) and two (C-D) basis trees. (A, C) column-wise sketch error and GW loss, (B, D) coefficient matrix. Orange boxes highlight basis trees. Magenta and teal boxes highlight trees with large and small sketch error/GW loss, respectively. Configuration: {\IFS} with MST.} 
    \label{fig:HC-matrices-2-3}
    \vspace{-2mm}
\end{figure}

Given 31 merge trees $\Tcal = \{T_0, \dots, T_{30}\}$ from the \emph{Heated Cylinder} dataset, we apply both CSS (specifically, {\IFS} and {\LSS}) and NMF to obtain a set of basis trees $\Scal$ and reconstruct the sketched trees. 
We first demonstrate that with only three basis trees, we  could obtain visually appealing sketched trees with small errors.  We then describe how the basis trees capture structural variations among the time-varying input. 	

\para{Sketched trees with {\IFS}}
We first illustrate our sketching results using {\IFS}. 
Based on our error analysis using the ``elbow method", three basis trees  appear to be the appropriate choice that strikes a balance between data summarization and structural preservation. 
The coefficient matrix, column-wise sketch error and GW loss (\autoref{fig:HC-matrices-2-3}) are used to guide our investigation into the quality of individual sketched trees. 
Trees with small GW losses or sketch errors are considered well sketched, whereas those with large errors are considered outliers. 
We give examples of a couple of well-sketched tree -- $T_7$ and $T_{15}$ (teal boxes) -- with several outliers -- $T_{8}$, $T_{24}$, and $T_{28}$ (magenta boxes) -- {\wrt} the chosen basis. 

As illustrated in \autoref{fig:teaser}, we compare a subset of input trees (B, blue boxes) against their sketched versions (C, red boxes).  
Even though we only use three basis trees, a large number of input trees -- such as $T_3$, $T_{15}$ -- and their sketched versions are indistinguishable with small errors.  
Even though $T_{8}$, $T_{24}$, and $T_{28}$ are considered outliers relative to other input trees, their sketched versions do not deviate significantly from the original trees. 
We highlight subtrees with noticeable structural differences before and after sketching in \autoref{fig:teaser}(C), whose roots are pointed by black arrows. 

\begin{figure}[!ht]
    \centering
    \includegraphics[width=0.8\columnwidth]{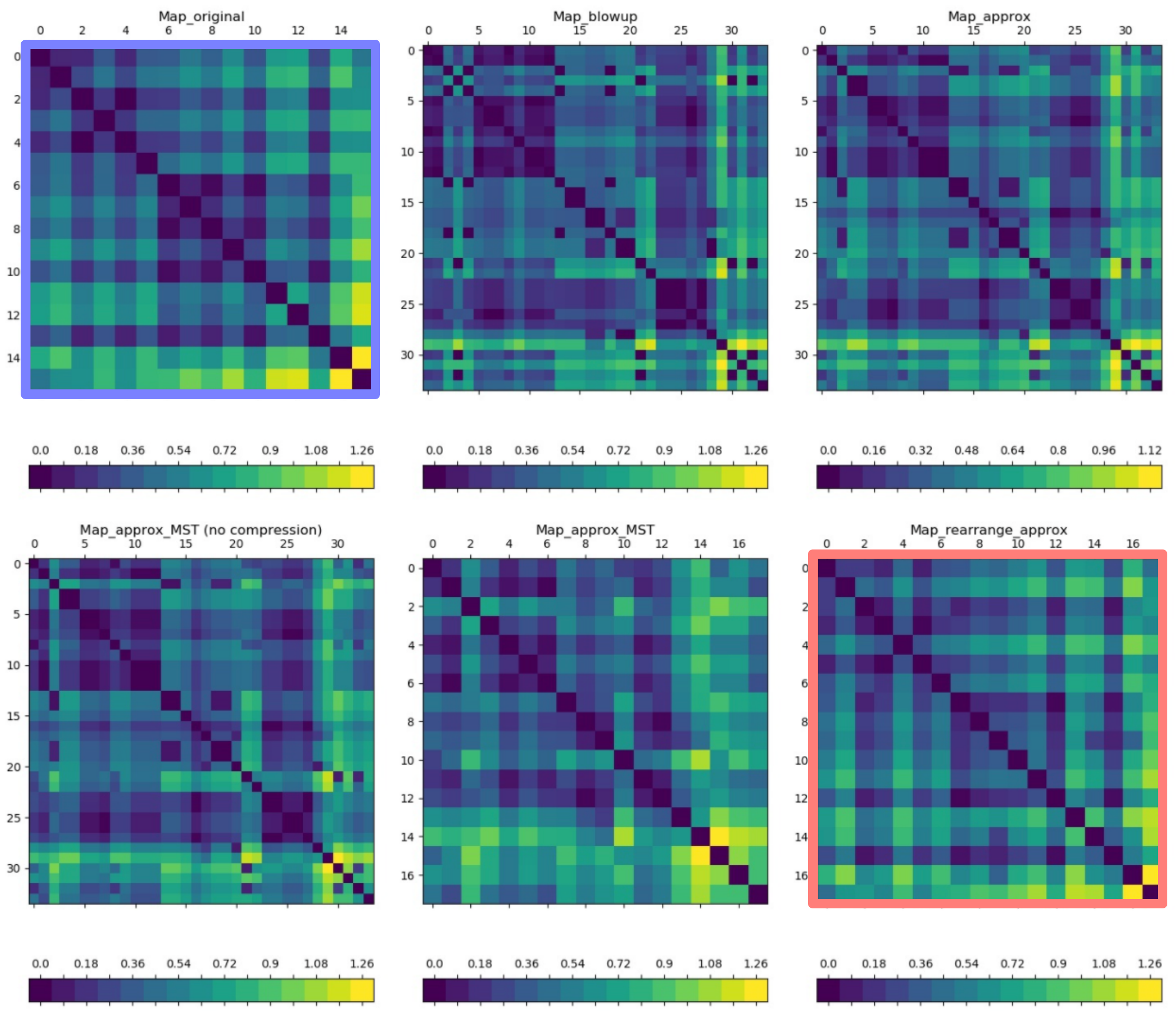}
    \vspace{-2mm}
    \caption{\emph{Heated Cylinder}: weight matrices associated with $T_{24}$ during the sketching process. Configuration: {\IFS} with MST.} 
    \label{fig:HC-T24}
\end{figure}

In \autoref{fig:HC-T24}, we further investigate the weight matrices from different stages of the sketching pipeline for tree $T = T_{24}$. 
From left to right, we show the weight matrix $W$ of the input tree, its blow-up matrix $W'$ (which is linearized to a column vector $a$), the approximated column vector $\hat{a}$ after sketching (reshaped into a square matrix), the weight matrix $\hat{W}'$ of the MST derived from the reshaped $\hat{a}$, the weight matrix of the MST after simplification, and root alignment $\hat{W}$ {\wrt} $T$. 
We observe minor changes between $W$ (blue box) and $\hat{W}$ (red box), which explain the structural differences before and after sketching in~\autoref{fig:teaser}. 

\begin{figure}[!ht]
    \centering
    \includegraphics[width=0.4\textwidth]{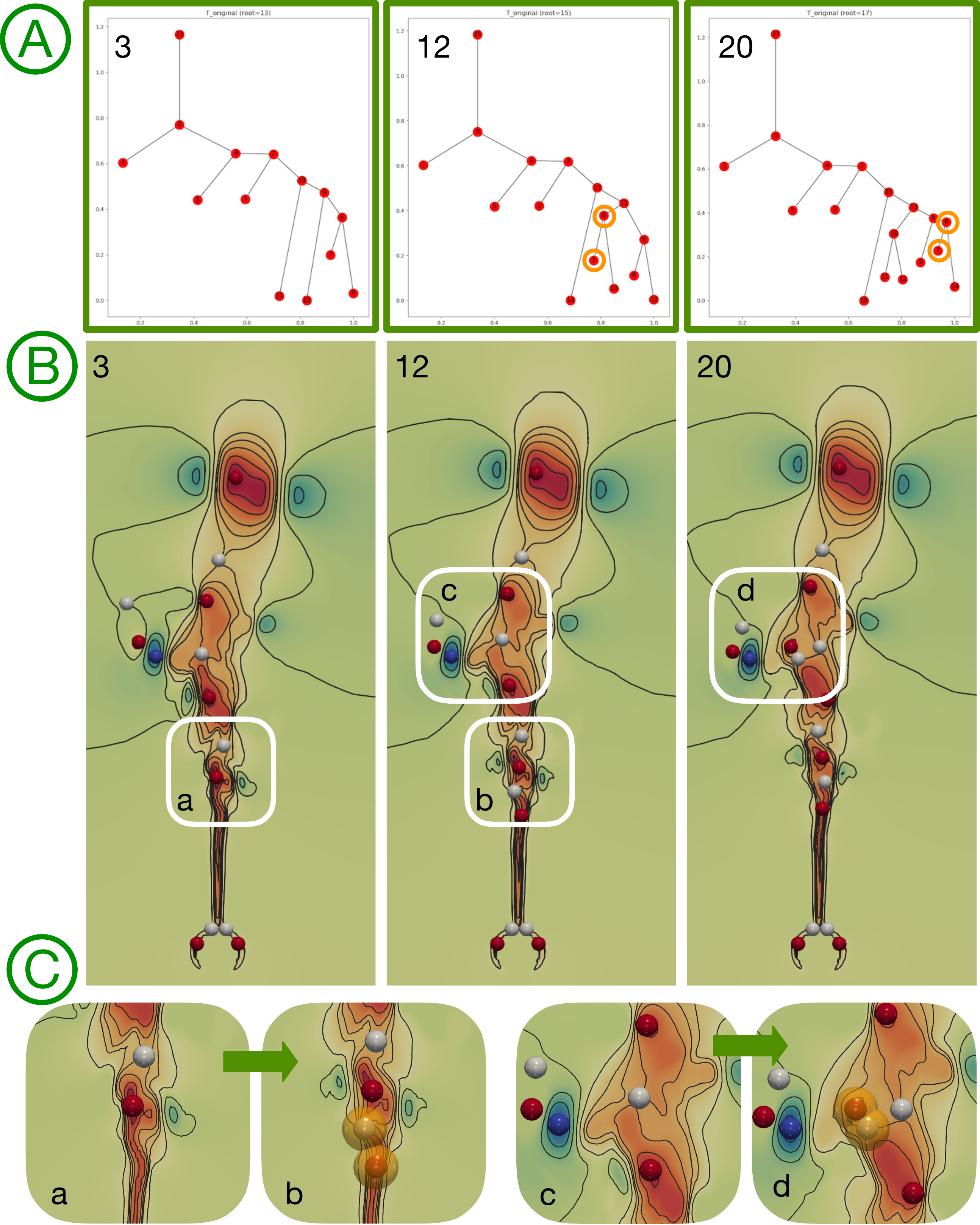}
    \caption{Sketching the \emph{Heated Cylinder} dataset with 3 basis trees: (A) basis trees where orange circles highlight topological changes {\wrt} nearby basis trees, (B) scalar fields that give rise to these basis trees, areas with critical points appearances/disappearances are shown with zoomed views in (C). Configuration: {\IFS} with MST.} 
    \label{fig:HC-basis}
\end{figure}

\para{Basis trees as representatives}
As shown in~\autoref{fig:HC-basis}(A), {\IFS} produces three basis trees, 
$\Scal = \{T_3, T_{12}, T_{20}\}$, which capture noticeable structural variations among the input merge trees. 
Specifically, moving from $T_3$ to $T_{12}$, and $T_{12}$ to $T_{20}$, a saddle-minima pair appears in the merge trees  respectively (highlighted by orange circles).
These changes in the basis trees reflect the appearances of critical points in the domain of the time-varying fields, see \autoref{fig:HC-basis}(B). 
In \autoref{fig:HC-basis}(C), we highlight (with orange balls) the appearances of these critical points in the domain. 

Furthermore, the coefficient matrix in \autoref{fig:HC-matrices-2-3}(B)   contains a number of yellow or light green blocks, indicating that consecutive input trees share similar coefficients {\wrt} the chosen basis and thus grouped together into clusters. 
Again, such a blocked structure indicates that the chosen basis trees appear to be good representatives of the clusters. 
In comparison, using just two basis trees does not capture the structural variations as well, where we see a slight degradation in the blocked structure thus sketching quality in \autoref{fig:HC-matrices-2-3}(C-D). 

\begin{figure}[!ht]
    \centering
    \includegraphics[width=0.8\columnwidth]{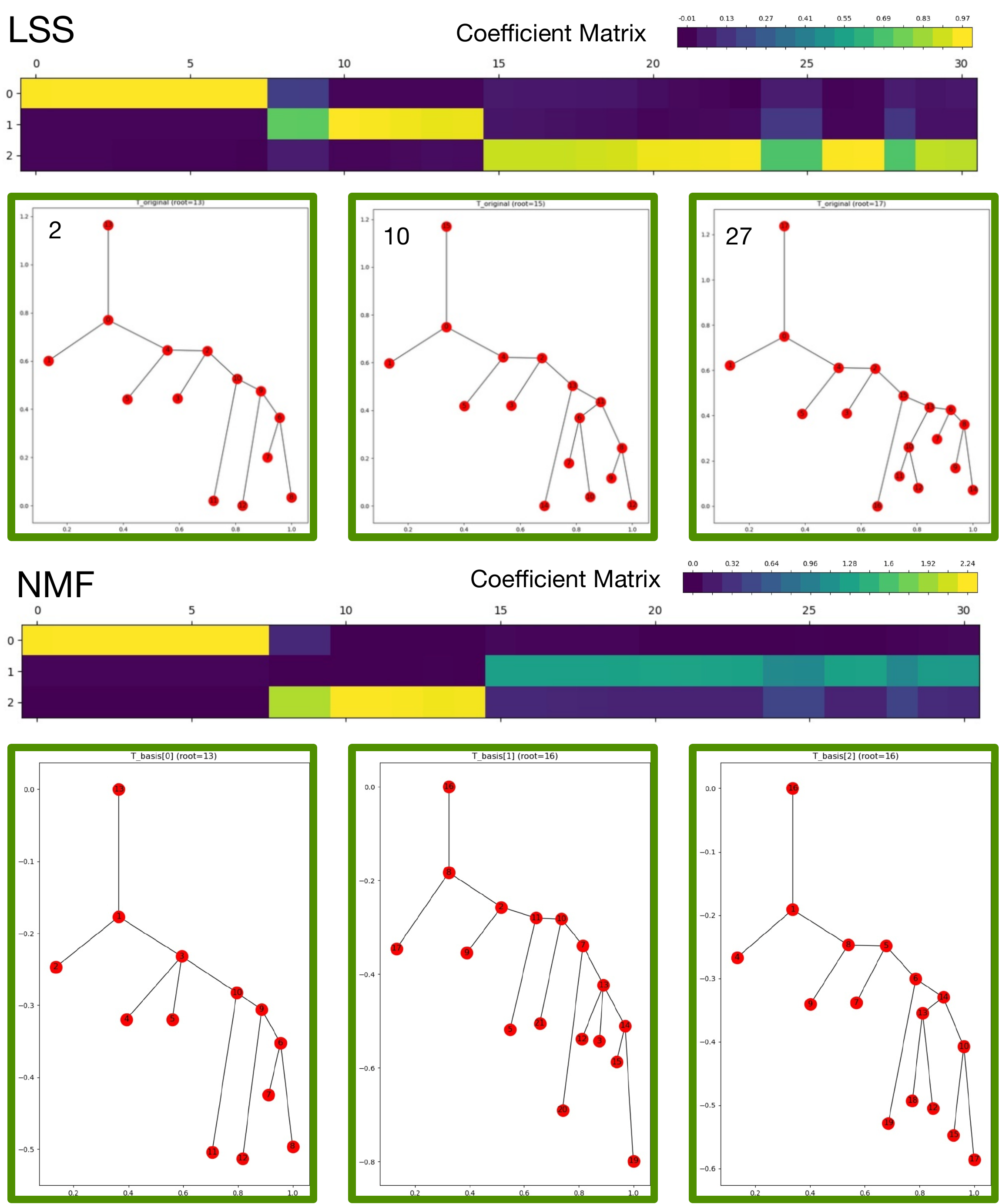}
    \caption{Coefficient matrices and basis trees used to sketch the \emph{Heated Cylinder} dataset with 3 basis trees with {\LSS} (top) and NMF (bottom).} 
    \label{fig:HC-LSS-NMF}
\end{figure}

\para{Sketching with {\LSS} and {\NMF}}
Additionally, we include the sketching results using {\LSS} and NMF as alternative strategies, again with three basis trees.  
{\LSS} gives basis trees $T_2, T_{10}$ and $T_{27}$ in \autoref{fig:HC-LSS-NMF} (Top), which are similar to the ones obtained by {\IFS} (\autoref{fig:HC-basis}).  
Using NMF, we show the three basis trees together with a coefficient matrix in~\autoref{fig:HC-LSS-NMF} (Bottom).  
Although these basis trees are generated by non-negative matrix factorization, that is, they do not correspond to any input trees; nevertheless they nicely pick up the structural variations in data and are shown to resemble the basis trees chosen by column selections. 
This shows that even through these matrix sketching techniques employ different (randomized) algorithms, they all give rise to reasonable choices of basis trees, which lead to good sketching results.

\subsection{Corner Flow Dataset}

The second dataset, referred to as the \emph{Cylinder Flow Around Corners} (\emph{Corner Flow} in short), arises from the simulation of a viscous 2D flow around two cylinders~\cite{BaezaRojoGunther2020, gerrisflowsolver}. 
The channel into which the fluid is injected is bounded by solid walls. 
A vortex street is initially formed at the lower left corner, which then evolves around the two corners of the bounding walls. 
We generate a set of merge trees from the magnitude of the velocity  fields of 100 time instances, which correspond to steps 801-900 from the original 1500 time steps. This dataset describes the formation of a one-sided vortex street on the upper right corner.

Given a set of $100$ merge trees from the \emph{Corner Flow} dataset, we first demonstrate that a set of $15$ basis trees chosen with {\IFS} gives visually appealing sketched trees with small error, based on the coefficient matrices and error analysis. 

\begin{figure}[!ht]
    \centering
    \includegraphics[width=1.0\columnwidth]{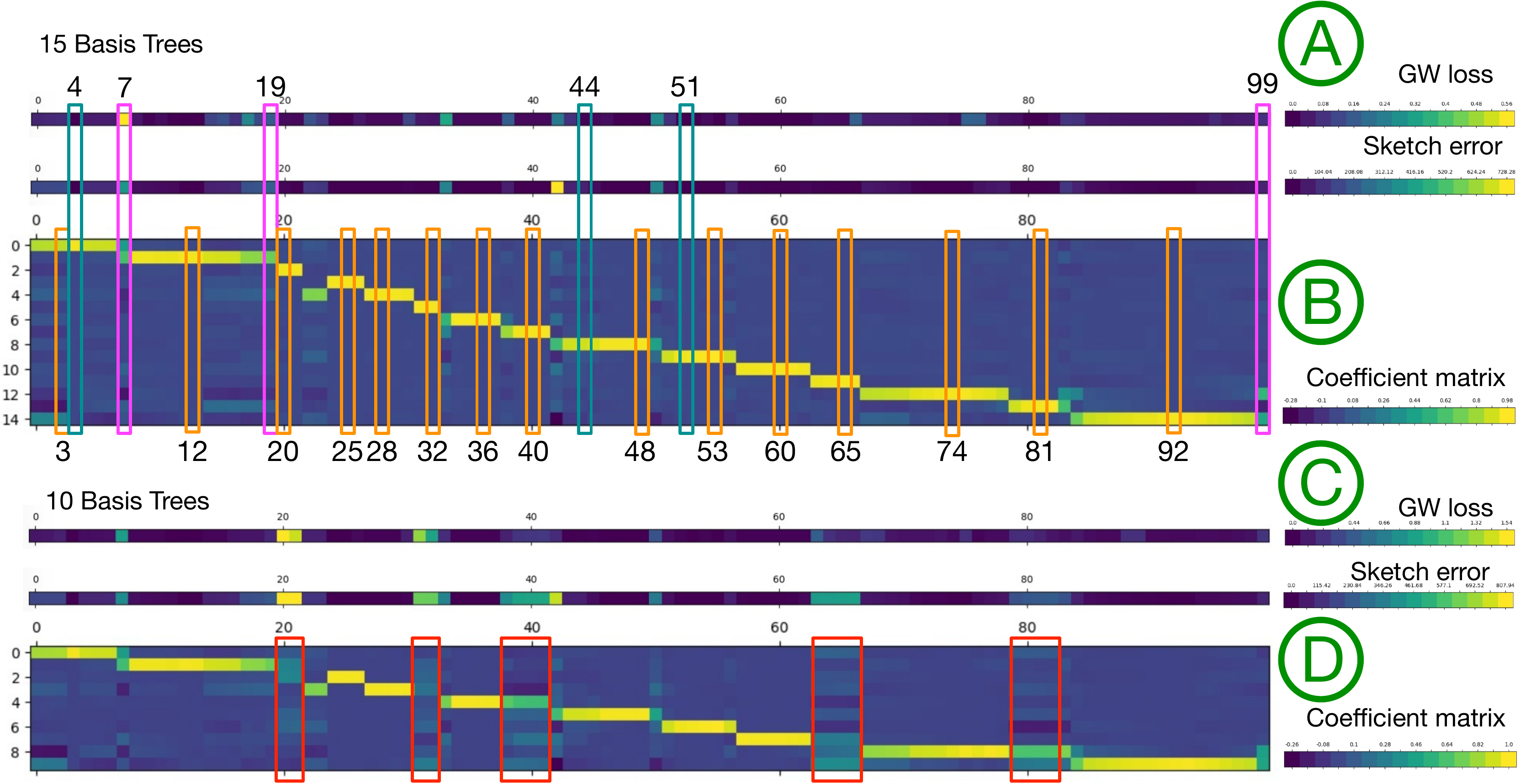}
    \caption{Sketching the \emph{Corner Flow} dataset with 15 (A, B) and 10 (C, D) basis trees. (A, C) column-wise sketch error and GW loss, (B, D) coefficient matrix. Orange boxes highlight basis trees. Magenta and teal boxes highlight trees with large and small sketch error/GW loss, respectively. Red boxes in (D) indicate trees that are better sketched with 15 basis trees. Configuration: {\IFS} with MST.}
    \label{fig:CF-matrices}
\end{figure}

\para{Coefficient matrices}
Using the ``elbow method" in the error analysis, we set $k=15$. 
We first compare the coefficient matrices generated using {\IFS}, for $k=10, 15$, respectively. 
Comparing~\autoref{fig:CF-matrices}(A) and (C), we see in general  improved  column-wise GW loss and sketch error using 15 instead of 10 basis trees. 
Furthermore, the coefficient matrix with 15 basis trees (B) contains better block structure than the one with 10 basis trees (D). 
Particularly, using additional basis trees improves upon the sketching results in regions enclosed by red boxes in (D).

\begin{figure*}[!ht]
    \centering
    \includegraphics[width=0.88\textwidth]{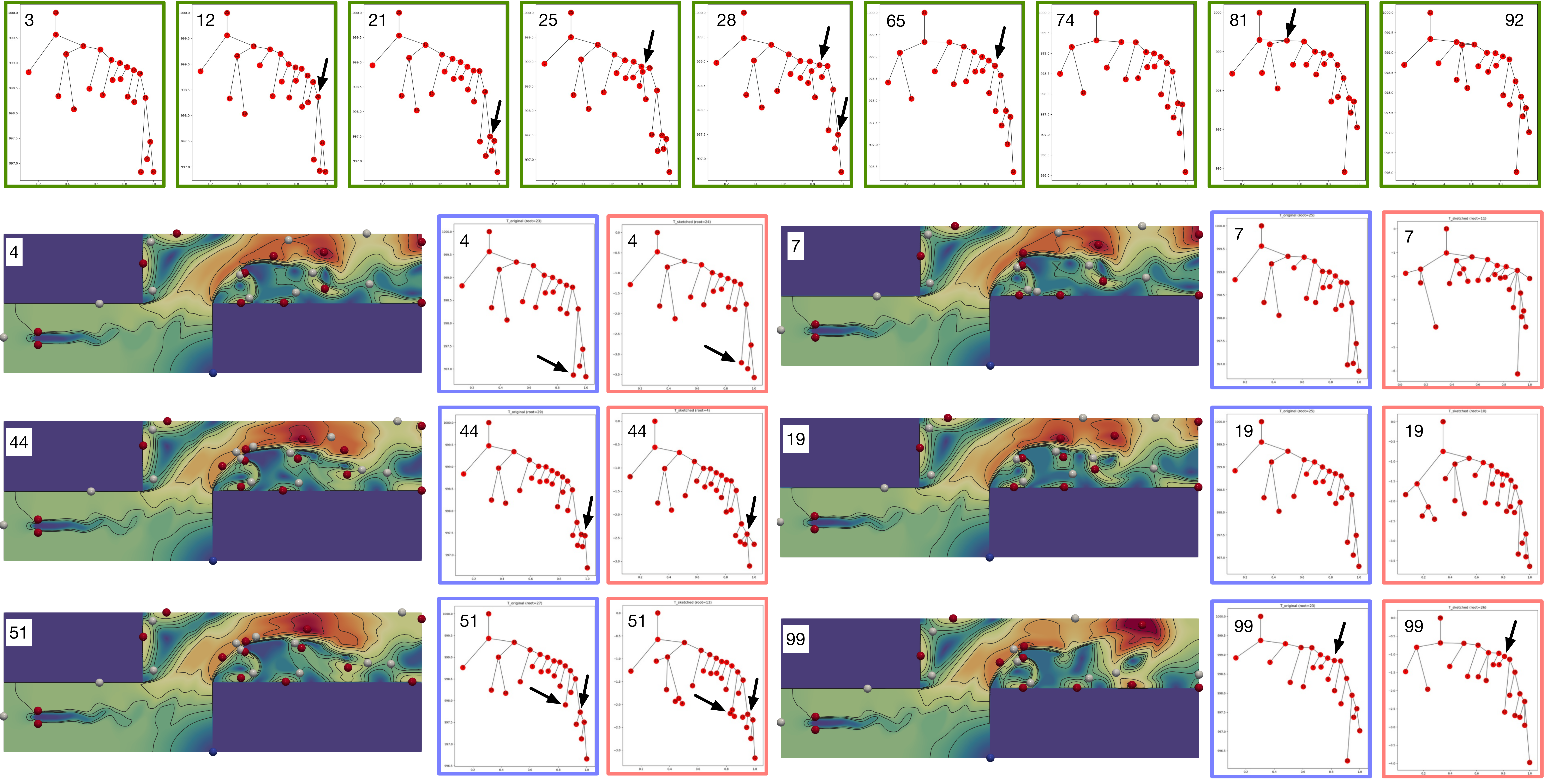}
    \vspace{-4mm}
    \caption{Individual sketched trees for the \emph{Corner Flow} dataset with 15 basis trees. Configuration: {\IFS} and MST. Green boxes are basis trees. Blue boxes are input trees while red boxes are sketched trees.} 
    \label{fig:CF-reps}
\end{figure*}

\para{Basis trees as representatives}
We thus report the sketching results with 15 basis trees under {\IFS}. 
The basis trees are selected with labels 3, 12, 21, 25, 28, 32, 36, 40, 48, 53, 60, 65, 74, 81, 92. 
Similar to the \emph{Heated Cylinder}, we observe noticeable structural changes among pairs of adjacent basis trees, which lead to a partition of the input trees into clusters with similar structures; see the block structure in~\autoref{fig:CF-matrices}(B).  
Thus the basis trees serve as good cluster representatives, as they are roughly selected one per block. 
We highlight the structural changes (with black arrows pointing at the roots of subtrees) among a subset of adjacent basis trees in~\autoref{fig:CF-reps} (green boxes).

 \begin{figure}[!ht]
    \centering
    \includegraphics[width=0.7\columnwidth]{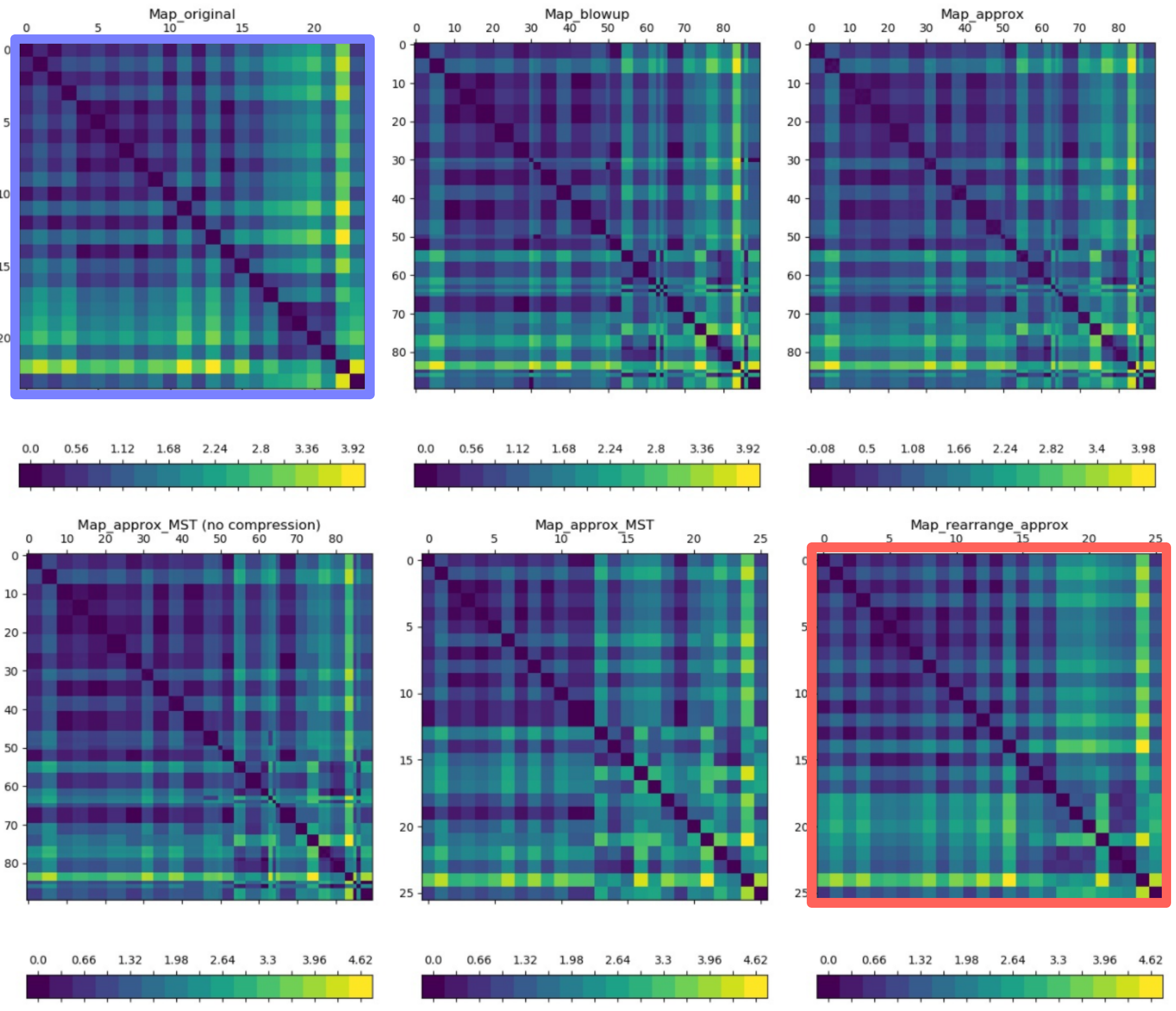}
    \vspace{-2mm}
    \caption{Sketching the \emph{Corner Flow} dataset with 15 basis trees. Weight matrices associated with $T_{99}$ during the sketching process. Configuration: {\IFS} with MST.} 
    \label{fig:HC-T99}
\end{figure}

\para{Sketched trees} 
Finally, we investigate individual sketched trees in~\autoref{fig:CF-reps}. 
We utilize the column-wise errors to select well-sketched trees (trees 4, 44, and 51) and outliers (trees 7, 19, and 99).
Trees with lower GW loss and sketch error are structurally similar to the chosen basis trees, and thus have a good approximation of their topology. 
For instance, the sketched tree 4 (red box) is almost indistinguishable {\wrt} to the original (blue box); the only difference is that the node pointed by the black arrow has a slightly higher function value. 

On the other hand, we observe that each outlier tree (e.g., $T_{7}, T_{99}$) is less visually appealing and has a higher sketch error.  
For instance, $T_{99}$ is shown to be a linear combination of two basis trees ($T_{74}$ and $T_{92}$), see also~\autoref{fig:CF-matrices}(B). 
 Its weight matrices before, during, and after sketching are shown in~\autoref{fig:HC-T99}, their differences before (blue box) and after (red box) sketching explain the observed structural discrepancies.

\subsection{Red Sea Dataset}

The third dataset, referred to as the \emph{Red Sea eddy simulation} (\emph{Red Sea} in short) dataset, originates from the IEEE Scientific Visualization Contest 2020\footnote{https://kaust-vislab.github.io/SciVis2020/}.  
The dataset is used to study the circulation dynamics and eddy activities of the Red Sea  (see~\cite{HoteitLuoBocquet2018, ZhanKrokosGuo2019, ZhanSubramanianYao2014}). 
For our analysis, we use merge trees that arise from velocity magnitude fields of an ensemble (named \emph{001.tgz}) with 60 times steps. 
Latter time steps capture the formation of various eddies, which are circular movements of water important for transporting energy and biogeochemical particles in the ocean. 

The \emph{Red Sea} dataset comes with 60 merge trees. 
The input does not exhibit natural clustering structures because many adjacent time instances give rise to trees with a large number of  structural changes. 
In this case, NMF performs better than {\IFS} and {\LSS} in providing visually appealing sketched trees since individual input trees do not capture these complex topological changes. 

\begin{figure*}[!ht]
    \centering
    \includegraphics[width=0.9\textwidth]{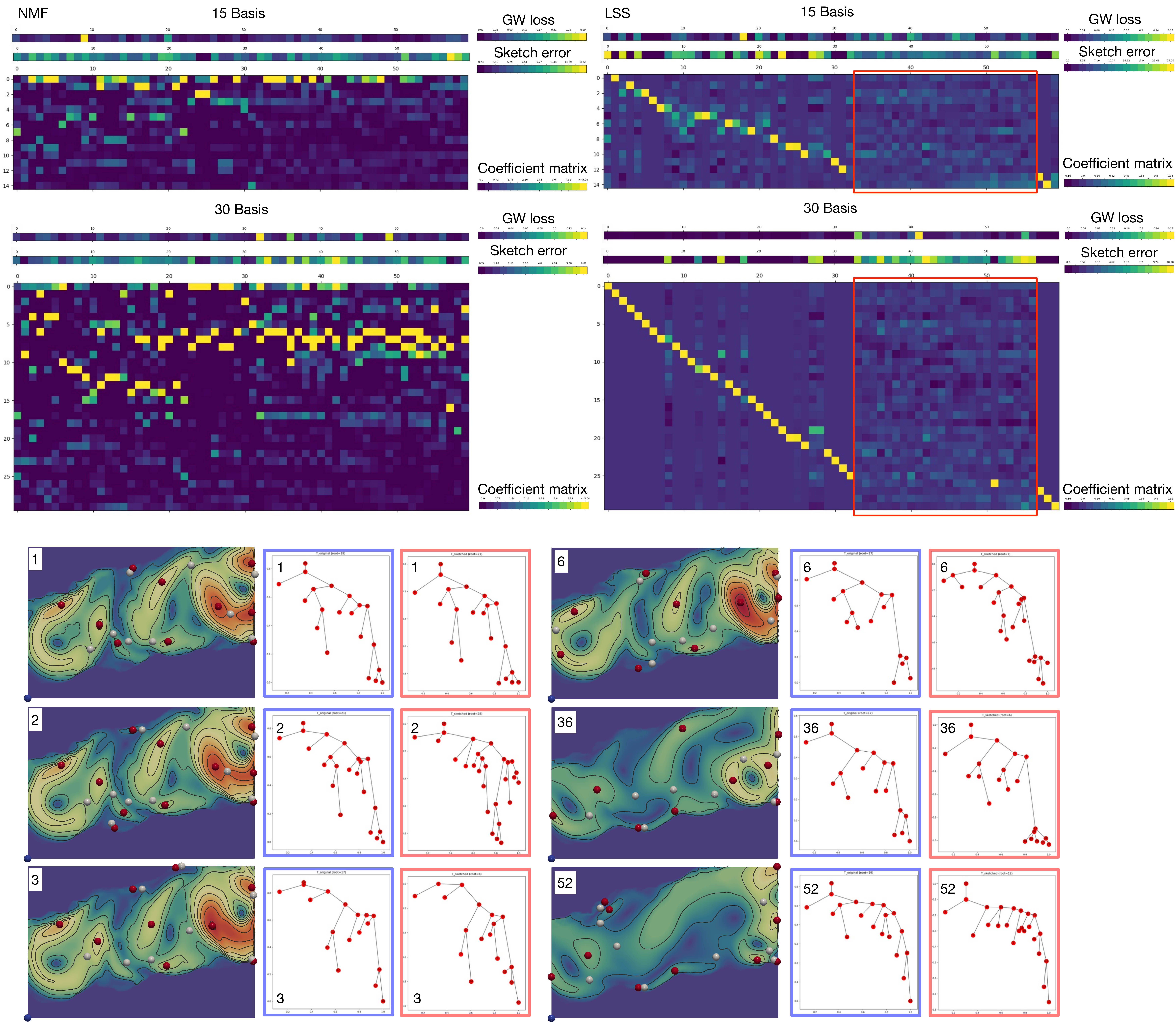}
     \vspace{-4mm}
    \caption{Top: Coefficient matrices, column-wise sketch error and GW loss, for sketching the \emph{Red Sea} dataset with 15 and 30 basis trees using NMF (left) and {\LSS} (right), respectively. Bottom: Individual sketched trees for the \emph{Red Sea} dataset together with their corresponding scalar fields. Configuration: 30 basis trees, NMF and MST. Blue boxes are input trees and red boxes are sketched trees.} 
    \label{fig:RS-reps}
\end{figure*}

\para{Coefficient matrices}
Using both NMF and {\LSS}, we compare the coefficient matrices for $k=15$ and $30$, respectively; see~\autoref{fig:RS-reps} (Top).
For {\LSS}, the input trees appear to have very diverse structures without clear large clusters. 
This phenomenon is evident by the lack of block structure (e.g., long  yellow rows) in the coefficient matrices. 
It is also interesting to notice that for LSS, there exists a subset of consecutive columns that contain few selected basis (e.g., red  boxes for $k=15, 30$).
On the other hand, using NMF, we obtain a slightly better block structure in the coefficient matrices.  
In general, the global sketch error and GW loss improve as we increase the number of basis.  

\para{Sketched trees}
In general, the \emph{Red Sea} dataset exhibits complex topological changes across time, thus it is not an easy dataset to sketch. 
We investigate the sketched individual trees with NMF and $k=30$ in~\autoref{fig:RS-reps} (Bottom). 
We visualize a number of sketched trees (trees labeled 1, 2, 3, 6, 36, 52) with varying errors together with their corresponding scalar fields. 
Given the diversity of the input trees, with 30 basis, we obtain a number of visually appealing sketched trees with minor structural differences (pointed by black arrows) {\wrt} to the original input trees. 
This show a great potential in using matrix factorization approaches to study and compress large collections of scientific datasets while preserving their underlying topology.

\section{Conclusion}
\label{sec:conclusion}

In this paper, we present a framework to sketch merge trees.
Given a set $\Tcal$ of merge trees of (possibly) different sizes, we compute a basis set of merge trees $\Scal$ such that each tree in $\Tcal$ can be approximately reconstructed using $\Scal$. 
We demonstrate the utility of our framework in sketching merge trees that arise from scientific simulations. 
Our framework can be used to obtain compact representations for downstream analysis and visualization, and to identify good representatives and outliers.
Our approach is flexible enough to be generalized to sketch other topological descriptors such as contour trees, Reeb graphs, and Morse--Smale graphs (e.g.,~\cite{CatanzaroCurryFasy2020}), which is left for future work.

\acknowledgments{
This work was partially funded by DOE DE-SC0021015. 
We thank Jeff Phillips for discussions involving column subset selection and Benwei Shi for his implementation on length squared sampling. We also thank Ofer Neiman for sharing the code on low stretch spanning tree.
}

\clearpage
\clearpage
\newpage


\clearpage
\newpage
\appendix
\section{Experimental Results with LSST}
\label{sec:lsst}

\begin{figure}[!ht]
    \centering
    \includegraphics[width=0.48\textwidth]{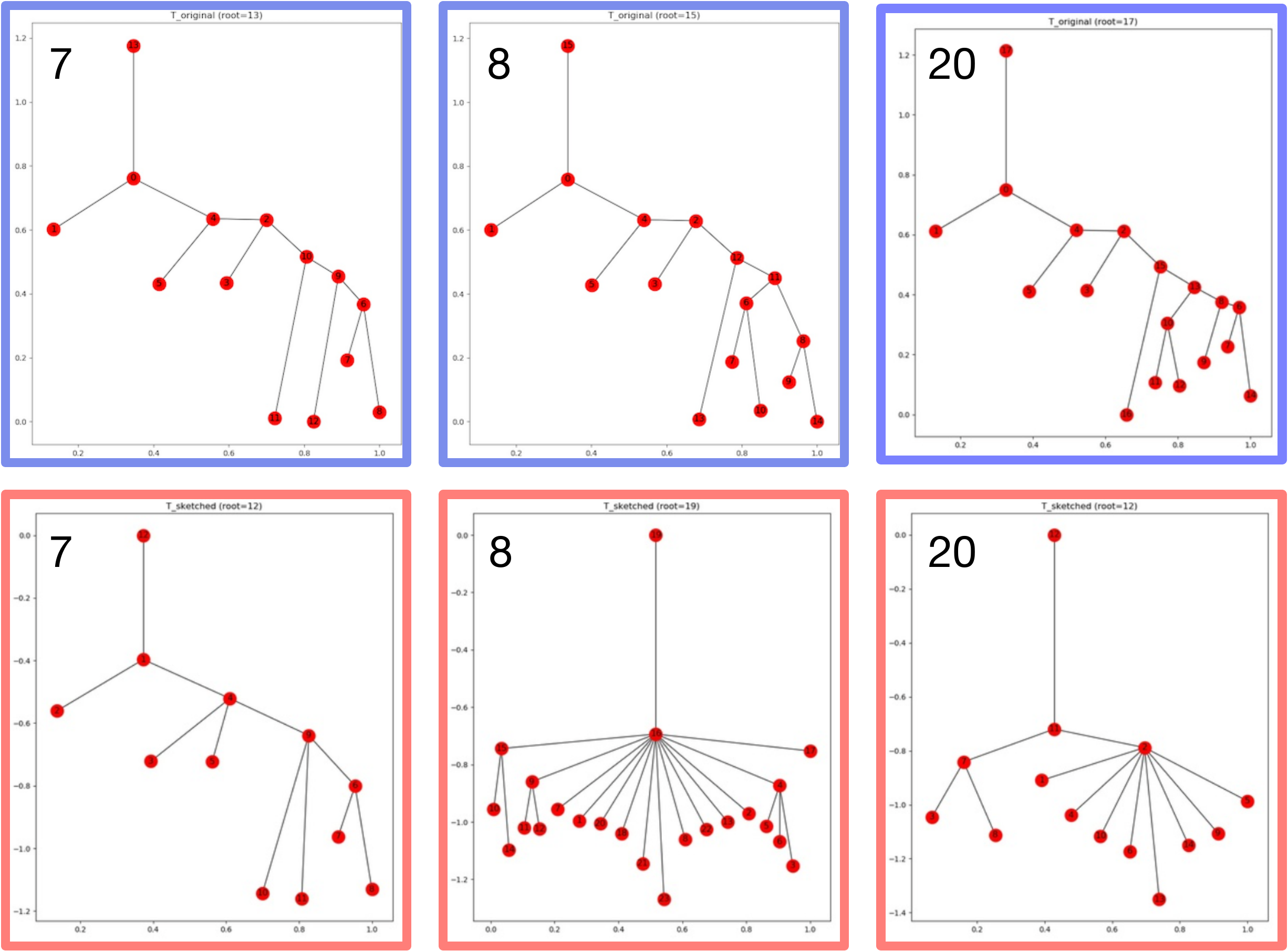}
    \caption{Sketched trees from the \emph{Heated Cylinder} dataset constructed with LSST based on {\IFS}.} 
    \label{fig:HC-IFS-LSST}
\end{figure}

For comparative purposes, we describe sketching results for \emph{Heated Cylinder} dataset using low-stretch spanning trees (LSST). 
The reconstructed sketched trees and basis trees using LSST are visually less appealing compared to the reconstruction using MST.
As shown in~\autoref{fig:HC-IFS-LSST}, the star-like features in the sketched trees are most likely a consequence of the petal decomposition algorithm of LSST~\cite{AbrahamNeiman2012}.

\begin{figure}[!ht]
    \centering
    \includegraphics[width=0.4\textwidth]{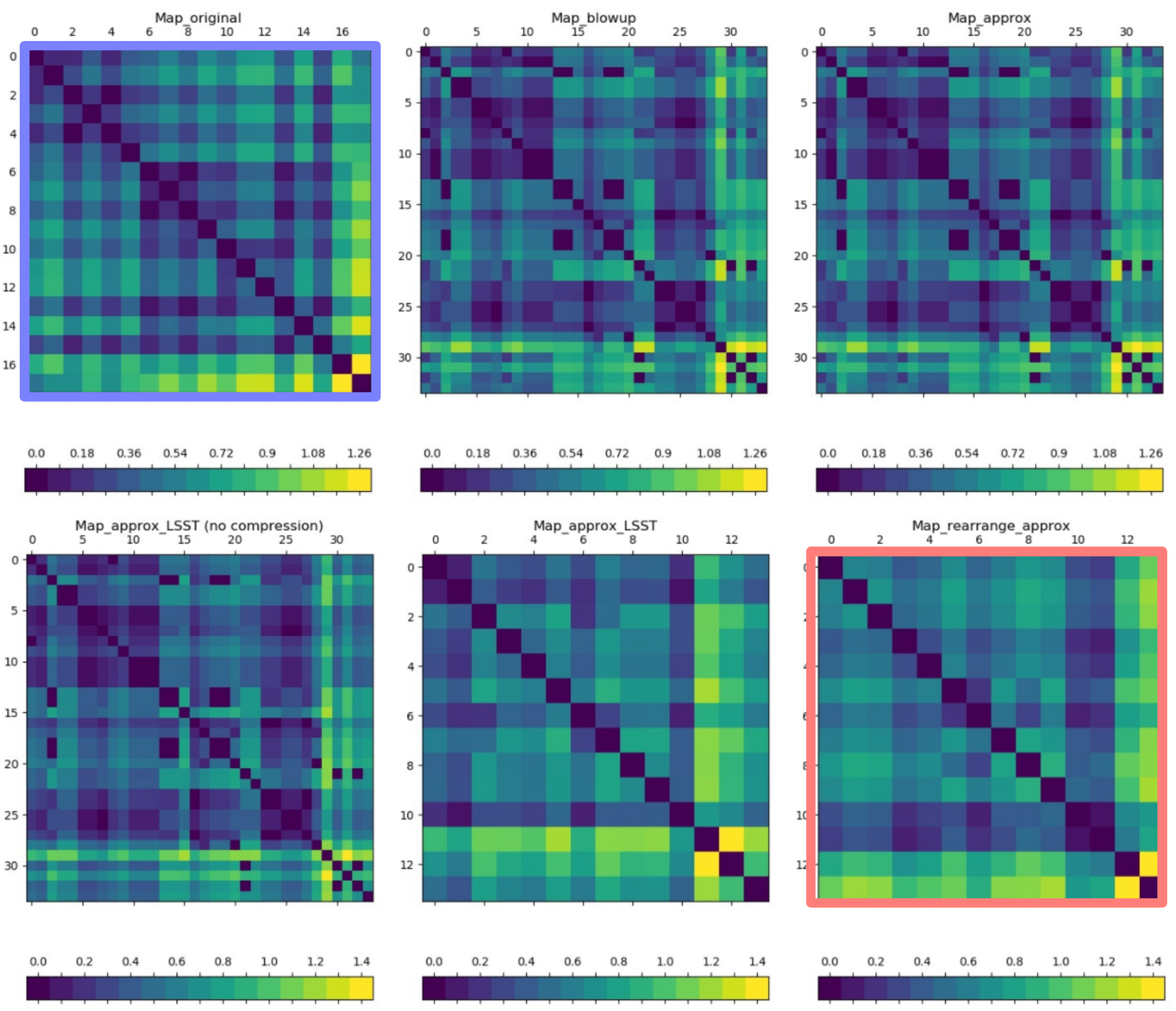}
        \caption{Sketching the \emph{Heated Cylinder} dataset with 3 basis trees. Weight matrices associated with $T_{20}$ during the sketching process. Configuration: {\IFS} with LSST.} 
    \label{fig:HC-IFS-LSST-weights}
\end{figure}

On the other hand, the weight matrices show that the LSST preserves some structures within the distance matrices, e.g.,~for $T_{20}$, before (blue box) and after (red box) sketching, see~\autoref{fig:HC-IFS-LSST-weights}.

\section{Theoretical Considerations}
\label{sec:theory}

We discuss some theoretical considerations in sketching merge trees. 
In the first two steps of our framework, we represent merge trees as metric measure networks and vectorize them via blow-up and alignment to a Fr{\'e}chet Mean using the GW framework~\cite{ChowdhuryNeedham2020}.
Each merge tree $T = (V, W, p) \in \Tcal$ is mapped to a column vector $a$ in matrix $A$, where $W$ captures the shortest path distances using function value differences as weights. 
The computation of the Fr{\'e}chet mean $\overline{T}$ is an optimization process, but the blow-up of $T$ and its alignment to $\overline{T}$ does not change the underlying distances between the tree nodes, which are encoded in $W$. 
Therefore, reshaping the column vector $a$ back to a pairwise distance matrix and computing its corresponding MST fully recovers the original input merge tree. 

In the third step, we sketch the matrix $A$ using either NMF or CSS.
Both matrix sketching techniques (albeit with different constraints) aim to obtain an approximation $\hat{A} = BY$ of $A$ that minimizes the error $\epsilon = \lVert A - \hat{A} \rVert_F$. 
Let $A_k$ denote the (unknown) best rank-$k$ approximation of $A$.  
In the case of CSS, the theoretical upper bound is given as a multiplicative error of the form $\epsilon \le \epsilon_k \cdot \lVert A - A_k\rVert_F$, where $\epsilon_k$ depends on the choice of $k$~\cite{DeshpandeVempala2006,BoutsidisMahoneyDrineas2009}, 
or it is given as an additive error $\epsilon \le \lVert A - A_k\rVert_F + \epsilon_{k, A}$, where $\epsilon_{k,A}$ depends on $k$ and $\lVert A \rVert_F$~\cite{DrineasKannanMahoney2006,MahoneDrineas2016}.
$\lVert A - A_k\rVert_F$ is often data dependent. 
In the case of NMF, a rigorous theoretical upper bound on $\epsilon$ remains unknown. 

Given an approximation $\hat{A}$ of $A$, the next step is to reconstruct a sketched merge tree from each column vector $\hat{a}$ of $\hat{A}$. 
We reshape $\hat{a}$ into an $n \times n$ matrix $\hat{W}$ and construct a sketched tree $\hat{T}$ by computing the MST or the LSST of $\hat{W}$.
The distance matrix $\hat{D}$ of the sketched tree $\hat{T}$ thus approximates the distance matrix $W'$ of the blow-up tree $T'$. 

When a sketched merge tree is obtained via a LSST, there is a theoretical upper bound on the relative distortion of the distances~\cite{AbrahamNeiman2012}, that is, $\theta \le O(\log n \log\log n)$ for 
$ \theta = \frac{1}{\binom{n}{2}} \sum_{x,x'} \left({\hat{D}(x, x')}/{\hat{W}(x, x')}\right).$
When a sketched merge tree is obtained via a MST, the theoretical bounds on $\lVert \hat{W} - \hat{D}\rVert_F$ are unknown, although, in practice, MST typically provides better sketched trees in comparison with LSST, as demonstrated in~\autoref{sec:results}.
Finally, although the smoothing process does not alter the tree structure significantly, it does introduce some error in the final sketched tree, whose theoretical bound  is not yet established. 

Therefore, while we have obtained good experimental results in sketching merge trees, there is still a gap between theory and practice for individual sketched trees. Filling such a gap is left for future work.  

\section{Implementation Details}
\label{sec:implementation}

In this section, we provide some implementation details for various algorithms employed in our merge tree sketching framework.

\para{Initializing the coupling probability distribution}
In \autoref{sec:method}, we introduce the blowup procedure that transforms a merge tree $T$ to a larger tree $T'$. 
This procedure optimizes the probability of coupling between $T$ and $\overline{T}$, the Fr\'{e}chet mean. 
Since the optimization process is finding a coupling matrix that is a local minimum of the loss function, similar input trees may give different coupling matrices due to the optimization process. 
This may affect the ordering of nodes in the blown-up trees, leading to completely different vectorization results and large sketch errors. 
Specifically for time-varying data, to ensure that adjacent trees are initialized with similar coupling probabilities {\wrt} $\overline{T}$, we use the  coupling probability between $T_{i-1}$ and $\overline{T}$ to initialize the coupling probability between $T_i$ and $\overline{T}$, for $1\leq i\leq N-1$. 
This strategy is based on the assumption that merge trees from adjacent time instances share similar structures.

\para{Matrix sketching algorithms}
We use two variants of column subset selection (CSS) algorithms, as well as non-negative matrix factorization (NMF) to sketch the data matrix $A$.
Here, we provide pseudocode for these matrix sketching algorithms.

\begin{itemize}\denselist
\item Modified Length Squared Sampling ({\LSS})
	\begin{enumerate}
	\item $s \leftarrow 0$, $B$ is an empty matrix, $A' = A$.  
	\item $s \leftarrow s+1$. Select column $c$ from $A'$ with the largest squared norm (or select $c$ randomly proportional to the squared norm) and add it as a column to $B$. Remove $c$ from $A'$. 
	\item For each remaining column $c'$ in $A'$ (\ie, $c' \neq c$), factor out the component along $c$ as:
	\begin{enumerate}
	      \item  $u \leftarrow c/\lVert c\rVert$  
	      \item  $c' \leftarrow   c' - \langle u,c' \rangle u$
	\end{enumerate}
	\item  While $s<k$, go to step 2. 
	\end{enumerate}
\end{itemize}

\begin{itemize}\denselist
\item Iterative Feature Selection ({\IFS})
	\begin{enumerate}
	\item Choose a subset of $k$ column indices $r = \{i_1, i_2, \dots, i_k\}$ uniformly at random.
	\item Construct subset $B_r = [a_{i_1}, a_{i_2}, \dots, a_{i_k}]$ of $A$ with columns indexed by $r$.
	\item Repeat for $j = 1, 2, \dots, k$:
		\begin{enumerate}
		      \item Let $X_{jl}$ denote matrix formed by replacing column $a_{i_j}$ with column $a_l$ in $B_r$, where $l\in [n]\setminus r$. Let $X_{jl}^{\pinv}$ denote its Moore-Penrose pseudoinverse.
		      \item Find $w = \argmin_{l\in [n]\setminus r} \lVert A - X_{jl} X_{jl}^{\pinv} A\rVert_F$.
		      \item $B_r \leftarrow X_{jw}$.
		      \item $r \leftarrow (r\setminus \{i_j\})\bigcup \{w\}$.
		\end{enumerate}
	\end{enumerate}
\end{itemize}

\begin{itemize}\denselist
\item Non-Negative Matrix Factorization ({\NMF})
	\begin{enumerate}
	\item Given $A$ and $k$, initialize $B\in \Rspace^{d\times k}$, $Y = X^T \in \Rspace^{k\times N}$ using the non-negative double singular value decomposition algorithm of Boutsidis and Gallopoulos~\cite{BoutsidisGallopoulos2008}.
	\item Normalize columns of $B$ and $X$ to unit $L_2$ norm. Let $E = A - BX^T$.
	\item Repeat until convergence: for $j = 1, 2, \dots, k$, 
		\begin{enumerate}
		    \item $Q \leftarrow E + b_j x_j^T$.
		    \item $x_j \leftarrow [Q^T b_j]_+$.
		    \item $b_j \leftarrow [Q x_j]_+$.
		    \item $b_j \leftarrow b_j / \lVert b_j \rVert$.
		    \item $E \leftarrow Q - b_j x_j^T$.
		\end{enumerate}
	\end{enumerate}
	Here, $[Q]_+$ means that all negative elements of the matrix $Q$ are set to zero.
\end{itemize}

\para{LSST algorithm}
We construct low stretch spanning trees (LSST) using the petal decomposition algorithm of Abraham and Neiman~\cite{AbrahamNeiman2012}.
Given a graph $G$, its LSST is constructed by recursively partitioning the graph into a series of clusters called \emph{petals}.
Each petal $P(x_0, t, r)$ is determined by three parameters: the center of the current cluster $x_0$, the target node of the petal $t$, and the radius of the petal $r$.

A cone $C(x_0, x, r)$ is the set of all nodes $v$ such that $d(x_0, x) + d(x, v) - d(x_0, v) \le r$.
A petal is defined as a union of cones of varying radii.
Suppose $x_0 \to x_1 \to \cdots \to x_k = t$ is the sequence of nodes on the shortest path between nodes $x_0$ and $t$.
Let $d_k$ denote the distance $d(x_k, t)$.
Then the petal $P(x_0, t, r)$ is defined as the union of cones $C(x_0, x_k, (r-d_k)/2)$ for all $x_k$ such that $d_k \le r$.

Beginning with a vertex $x_0$ specified by the user, the algorithm partitions the graph into a series of petals.
When no more petals can be obtained, all the remaining nodes are put into a cluster called the stigma.
A tree structure, rooted in the stigma, is constructed by connecting the petals and the stigma using some of the intercluster edges.
All other edges between clusters are dropped.
This process is applied recursively within each petal (and the stigma) to obtain a spanning tree structure.

\para{Merge tree simplification}
To reconstruct a sketched tree, we reshape the sketched column vector $\hat{a}$ of $\hat{A}$ into an $n\times n$ matrix $\hat{W'}$, and obtain a tree structure $\hat{T'}$ by computing its MST or LSST.
$\hat{T'}$ is an approximation of the blown-up tree $T'$.
To get a tree approximation closer to the original input tree $T$, we further simplify $\hat{T'}$ as described below.  

The simplification process has two parameters. 
The first parameter $\alpha$ is used to merge internal nodes that are too close ($\le\alpha$) to each other.
Let $R$ be the diameter of $\hat{T}'$ and $n$ the number of nodes in $\hat{T}'$. 
$\alpha$ is set to be $c_{\alpha}R/n^2$ for $c_{\alpha} \in \{0.5, 1, 2\}$. 
A similar parameter was used in simplifying LSST in~\cite{AbrahamNeiman2012}. 
The second parameter $\beta = c_{\beta} R/n$ is used to merge leaf nodes that are too close ($\leq \beta$) to the parent node, where $c_{\beta} \in \{0.5, 1, 2\}$. Let $\hat{W}'$ be the weight matrix of $\hat{T}'$. The simplification process is as follows:  
\begin{enumerate}\denselist
    \item Remove from $\hat{T'}$ all edges $(u, v)$ where $\hat{W}'(u, v) \leq  {\alpha}$.
    \item Merge all leaf nodes $u$ with their respective parent node $v$ if $\hat{W}'(u, v) \leq \beta$.
    \item Remove all the internal nodes.
\end{enumerate}
The tree $\hat{T}$ obtained after simplification is the final sketched tree. 

\para{Merge tree layout}
To visualize both input merge trees and sketched merge trees, we experiment with a few strategies. 
To draw an input merge tree $T$ equipped with a function defined on its nodes, $f: V \to \Rspace$, we set each node $u \in V$ to be at location $(x_u, y_u)$; where $y_u = f(u)$, and $x_u$ is chosen within a bounding box while avoiding edge intersections. 
The edge $(u,v)$ is drawn proportional to its weight $W(u,v)=|f(u)-f(v)|=|y_u-y_v|$.

To draw a sketched tree as a merge tree, we perform the following steps: 
\begin{enumerate}\denselist
    \item Fix the root of the sketched tree at $(0, 0)$.
    \item The y-coordinate of each child node is determined by the weight of the edge between the node and its parent.
    \item The x-coordinate is determined by the left-to-right ordering of the child nodes. We consider to order the child nodes that share the same parent node by using a heuristic strategy described below.
    \begin{enumerate}
    \item Sort the child nodes by their size of the subtrees of which the child node is the root in ascending order. This is trying to keep larger subtrees on the right so the overall shape of the tree is protected and straightforward to read. 
    \item If the sizes of multiple subtrees are the same, we apply the following strategy: we sort child nodes by their distances to the parent node in descending order. Suppose the order of child nodes after sorting is $c_1, c_2, \dots, c_t$. If $t$ is odd, we reorder the nodes from left to right as $c_t, c_{t-2}, c_{t-4}, \dots, c_3, c_1, c_2, c_4, \dots, c_{t-3}, c_{t-1}$. If $t$ is even, we reorder the nodes as $c_{t-1}, c_{t-3}, c_{t-5}, \dots, c_3, c_1, c_2, c_4, \dots, c_{t-2}, c_{t}$. 
\end{enumerate}
\end{enumerate}
\noindent The idea is to keep the child nodes that have a larger distance to the parent near the center to avoid edge crossings between sibling nodes and their subtrees.

Our layout strategy assumes that the trees are rooted.
However, $\hat{T}$, which is our approximation of $T$, is not rooted.
In our experiments, we use two different strategies to pick a root for $\hat{T}$ and align $T$ and $\hat{T}$ for visual comparison. 

Using the \textbf{balanced layout} strategy, we pick the node $u$ of $\hat{T}$ that minimizes the sum of distances to all other nodes. Set $u$ to be the \emph{balanced root} of $\hat{T}$. Similarly, we find the balanced root $v$ of the input tree $T$. $T$ and $\hat{T}$ are drawn using the balanced roots. 

Using the \textbf{root alignment} strategy, we obtain the root node of the sketched tree by keeping track of the root node during the entire sketching process. We can get the root node of $T'$ because it is either a duplicate node or the same node of the root node in $T$. Then we can get the root node in $\hat{T}'$, as the labels in the sketched blown-up tree are identical to $T'$. Lastly, by keeping track of the process of merge tree simplification, we can know the label of the root of $\hat{T}$.

\para{Other implementation details}
Our framework is mainly implemented in \textsf{Python}. 
The code to compute LSST and MST from a given weight matrix is implemented in \textsf{Java}. 
For data processing and merge tree visualization, we use \textsf{Python} packages, including \textsf{numpy}, \textsf{matplotlib}, and \textsf{networkx}. 
In addition, the GW framework of Chowdhury and Needham~\cite{ChowdhuryNeedham2020} requires the \textsf{Python Optimal Transport (POT)} package.

\section{Detailed Error Analysis}
\label{sec:error-analysis}

In~\autoref{fig:elbow}, we see that all global sketch errors and most global GW losses decrease as the number of basis trees $k$ increases. 
The decrease of global sketch errors is not surprising as this is a direct consequence of matrix sketching when $k$ increases. 
For the \emph{Heated Cylinder} dataset, we see that the global GW loss does not decrease drastically for $k\geq3$.   
This is because that almost all input trees are well sketched at $k=3$, resulting in small column-wise GW losses. 
This is the intuition behind our ``elbow method". 

In terms of the global sketch error, {\LSS} strategy appear to have the worst performance when $k$ is small, while {\NMF} consistently performs the best for the \emph{Red Sea} dataset with the most complicated topological variations.  
{\IFS} and {\LSS} overall have similar performances in all datasets, while {\IFS} usually performs slightly better than {\LSS} when $k$ is small.
We report below the exact errors for a single run (with a fixed seed for the randomization) across increasing $k$ values for the three datasets. 
We compare across three sketching techniques, {\NMF}, {\LSS}, and {\IFS}. 
We compare GW losses obtained using both MST and LSST strategy. 

\begin{table}[]
\resizebox{0.98\columnwidth}{!}{
\begin{tabular}{ccccc}
\multirow{2}{*}{Sketching Method} & \multirow{2}{*}{$k$} & \multicolumn{2}{c}{GW loss} & \multirow{2}{*}{Sketch error} \\
                                  &                               & MST          & LSST         &                              \\ \hline
\multirow{4}{*}{{\NMF}}              & 2                             & 1.4435       & 0.5233       & 123.1844                     \\
                                  & 3                             & 0.3420       & 0.5138       & 38.1245                      \\
                                  & 4                             & 0.4892       & 0.4992       & 27.7735                      \\
                                  & 5                             & 0.2927       & 0.3906       & 12.1199                      \\ \hline
\multirow{4}{*}{{\IFS}}          & 2                             & 0.4581       & 0.7500       & 156.4803                     \\
                                  & 3                             & 0.1756       & 0.5422       & 46.8917                      \\
                                  & 4                             & 0.1514       & 0.5453       & 28.1174                      \\
                                  & 5                             & 0.1159       & 0.4944       & 10.0892                      \\ \hline
\multirow{4}{*}{{\LSS}}          & 2                             & 1.7292       & 0.6311       & 247.4370                     \\
                                  & 3                             & 0.1574       & 0.6258       & 52.3285                      \\
                                  & 4                             & 0.1157       & 0.6743       & 34.1059                      \\
                                  & 5                             & 0.1416       & 0.5837       & 14.4842                      \\ \hline
\end{tabular}}
\\
\caption{GW losses and sketch errors of sketching the \textit{Heated Cylinder} dataset with increasing $k$.}
\label{tab:heated_cylinder_loss}
\end{table}

\para{\emph{Heated Cylinder}}
In \autoref{tab:heated_cylinder_loss}, we see that global GW losses with MST become stable for $k\geq3$. However, the global GW losses with LSST do  not converge as $k$ increases to $5$. 
This is partially due to the fact that LSST does not recover the merge tree as well as the MST. 
Among three sketching methods, at $k=3$, {\IFS} and {\LSS} have better performance than {\NMF} for $k\geq3$ {\wrt} the GW loss, while {\NMF} has the best performance on the sketch error.

\begin{table}[]
\resizebox{0.98\columnwidth}{!}{
\begin{tabular}{ccccc}
\multirow{2}{*}{Sketching Method} & \multirow{2}{*}{$k$} & \multicolumn{2}{c}{GW loss} & \multirow{2}{*}{Sketch error} \\
                                  &                               & MST          & LSST         &                              \\ \hline
\multirow{6}{*}{{\NMF}}              & 5                             & 46.0725      & 16.6348      & 23217.9002                   \\
                                  & 10                            & 27.2971      & 13.3785      & 10359.9521                   \\
                                  & 15                            & 21.2988      & 12.4028      & 5119.7659                    \\
                                  & 20                            & 14.7962      & 12.2321      & 2724.0766                    \\
                                  & 25                            & 13.3384      & 11.9885      & 1781.4510                    \\
                                  & 30                            & 13.2442      & 12.0104      & 1189.0048                    \\ \hline
\multirow{6}{*}{{\IFS}}          & 5                             & 37.9071      & 16.9597      & 29547.6731                   \\
                                  & 10                            & 12.8413      & 13.2241      & 12332.1347                   \\
                                  & 15                            & 6.5156       & 12.7953      & 4780.1059                    \\
                                  & 20                            & 4.5175       & 9.4656       & 2023.8101                    \\
                                  & 25                            & 3.3680       & 7.5935       & 888.4257                     \\
                                  & 30                            & 2.9452       & 5.9000       & 412.4219                     \\ \hline
\multirow{6}{*}{{\LSS}}          & 5                             & 71.8523      & 19.8988      & 39499.2563                   \\
                                  & 10                            & 33.5862      & 15.0740      & 21475.1222                   \\
                                  & 15                            & 8.6099       & 11.5266      & 8254.5006                    \\
                                  & 20                            & 5.5886       & 9.2652       & 3041.7406                    \\
                                  & 25                            & 4.0473       & 8.2447       & 1280.3685                    \\
                                  & 30                            & 2.9364       & 9.4942       & 725.1093                     \\ \hline
\end{tabular}}
\\
\caption{GW losses and sketch errors of sketching the \textit{Corner Flow} dataset with increasing $k$.}
\label{tab:corner_flow_loss}
\end{table}

\para{\emph{Corner Flow}}
In \autoref{tab:corner_flow_loss}, sketch errors decrease drastically as $k$ increases. 
For the two CSS sketching methods -- {\IFS} and {\LSS} -- the ``elbow" point of the GW loss with MST is at $k = 15$. 
Therefore we report our results using $k=15$ basis trees. 
Based on the performance of GW loss for $k\geq15$, {\IFS} performs the best among the three sketching methods.

\begin{table}[]
\resizebox{0.98\columnwidth}{!}{
\begin{tabular}{ccccc}
\multirow{2}{*}{Sketching Method} & \multirow{2}{*}{$k$} & \multicolumn{2}{c}{GW loss} & \multirow{2}{*}{Sketch error} \\
                                  &                               & MST          & LSST         &                              \\ \hline
\multirow{6}{*}{{\NMF}}              & 5                             & 6.9777       & 0.6956       & 933.3919                     \\
                                  & 10                            & 4.9383       & 0.5673       & 627.3957                     \\
                                  & 15                            & 2.8533       & 0.5011       & 440.2140                     \\
                                  & 20                            & 2.5431       & 0.5047       & 329.9397                     \\
                                  & 25                            & 1.5934       & 0.4128       & 246.5278                     \\
                                  & 30                            & 1.3152       & 0.3717       & 183.8884                     \\ \hline
\multirow{6}{*}{{\IFS}}          & 5                             & 3.7679       & 0.7960       & 1230.5378                    \\
                                  & 10                            & 2.1023       & 0.5802       & 830.7574                     \\
                                  & 15                            & 2.3979       & 0.5181       & 553.0398                     \\
                                  & 20                            & 1.4692       & 0.3485       & 397.0191                     \\
                                  & 25                            & 1.0142       & 0.2724       & 283.8923                     \\
                                  & 30                            & 0.7326       & 0.2579       & 201.8570                     \\ \hline
\multirow{6}{*}{{\LSS}}          & 5                             & 3.8604       & 0.8583       & 1629.3500                    \\
                                  & 10                            & 2.2035       & 0.5904       & 910.2818                     \\
                                  & 15                            & 2.3660       & 0.4774       & 603.2064                     \\
                                  & 20                            & 1.6389       & 0.4137       & 437.7015                     \\
                                  & 25                            & 1.1700       & 0.3275       & 313.6370                     \\
                                  & 30                            & 0.8721       & 0.3665       & 220.9125                     \\ \hline
\end{tabular}}
\\
\caption{GW losses and sketch errors of sketching the \textit{Red Sea} dataset with increasing $k$.}
\label{tab:red_sea_loss}
\end{table}

\para{\emph{Red Sea}}
In \autoref{tab:red_sea_loss}, the ``elbow" point is not as obvious. 
This is because the \emph{Red Sea} dataset is the hardest to sketch, as the  input trees contain significantly diverse topological structures.  
Similar to the results of other two datasets, {\IFS} has overall the best performance {\wrt} the GW loss, {\LSS} the second, and {\NMF} the worst.

\para{Average performance across multiple runs}
Our pipeline includes randomization to set initial states, including {\NMF}, {\IFS}, and the LSST algorithm. 
Therefore, we report the average GW loss and sketch error across 10 runs, each with a distinct random seed. 
By comparing \autoref{tab:multiple_runs_loss} with \autoref{tab:heated_cylinder_loss}, \autoref{tab:corner_flow_loss}, and \autoref{tab:red_sea_loss}, we see that the average GW losses and sketch errors across 10 different runs are close to the results of a single run with a fixed random seed. This shows that our pipeline has a reasonably stable performance on sketching merge trees. 

\begin{table}[]
\resizebox{0.98\columnwidth}{!}{
\begin{tabular}{c|ccccc}
\multirow{2}{*}{Dataset}         & \multirow{2}{*}{Sketching Method} & \multirow{2}{*}{$k$} & \multicolumn{2}{c}{Avg. GW loss} & \multirow{2}{*}{Avg. Sketch error} \\
                                 &                                   &                               & MST             & LSST           &                                   \\ \hline
\multirow{6}{*}{\emph{Heated Cylinder}} & \multirow{2}{*}{{\NMF}}              & 3                             & 0.3420          & 0.4825         & 38.1245                           \\
                                 &                                   & 5                             & 0.2927          & 0.4247         & 12.1199                           \\ \cline{2-6} 
                                 & \multirow{2}{*}{{\IFS}}          & 3                             & 0.1756          & 0.5981         & 46.8917                           \\
                                 &                                   & 5                             & 0.1159          & 0.4736         & 10.0892                           \\ \cline{2-6} 
                                 & \multirow{2}{*}{{\LSS}}          & 3                             & 0.1574          & 0.7299         & 52.3285                           \\
                                 &                                   & 5                             & 0.1416          & 0.5152         & 14.4842                           \\ \hline
\multirow{6}{*}{\emph{Corner Flow}}     & \multirow{2}{*}{{\NMF}}              & 15                            & 21.9481         & 12.8625        & 5119.6605                         \\
                                 &                                   & 30                            & 13.2744         & 12.1794        & 1188.9742                         \\ \cline{2-6} 
                                 & \multirow{2}{*}{{\IFS}}          & 15                            & 6.5156          & 12.0558        & 4780.1059                         \\
                                 &                                   & 30                            & 2.9452          & 6.0414         & 412.4219                          \\ \cline{2-6} 
                                 & \multirow{2}{*}{{\LSS}}          & 15                            & 8.6099          & 11.0902        & 8254.5006                         \\
                                 &                                   & 30                            & 2.9364          & 9.1941         & 725.1093                          \\ \hline
\multirow{6}{*}{\emph{Red Sea}}         & \multirow{2}{*}{{\NMF}}              & 15                            & 3.2322          & 0.4936         & 442.6564                          \\
                                 &                                   & 30                            & 1.3653          & 0.4398         & 188.4722                          \\ \cline{2-6} 
                                 & \multirow{2}{*}{{\IFS}}          & 15                            & 2.3979          & 0.5362         & 553.0398                          \\
                                 &                                   & 30                            & 0.7600          & 0.2706         & 201.8789                          \\ \cline{2-6} 
                                 & \multirow{2}{*}{{\LSS}}          & 15                            & 2.3660          & 0.5181         & 603.2064                          \\
                                 &                                   & 30                            & 0.8721          & 0.3512         & 220.9125                          \\ \hline
\end{tabular}}
\\
\caption{Average GW loss and Sketch error of 10 different runs for the sketching algorithms.}
\label{tab:multiple_runs_loss}
\end{table}

\end{document}